\documentclass[10pt]{emulateapj}

\usepackage{amsmath}
\usepackage{graphicx}
\usepackage{natbib}
\usepackage{amssymb}

\def\tint{T_{\rm int}}

\def\teq{T_{\rm eq}}

\begin{document}

\title{Planetary population synthesis coupled with atmospheric escape: a statistical view of evaporation}

\author{Sheng Jin\altaffilmark{1,2,3}, Christoph Mordasini$^{\star}$\altaffilmark{2}, Vivien Parmentier\altaffilmark{4}, \\ Roy van Boekel\altaffilmark{2}, Thomas Henning\altaffilmark{2}, Jianghui Ji\altaffilmark{1,5}}

\affil{$^1$
Purple Mountain Observatory, Chinese Academy of Sciences, Nanjing 210008, China; shengjin@pmo.ac.cn
}

\affil{$^2$
Max-Planck-Institut f\"ur Astronomie, K\"onigstuhl 17, D-69117 Heidelberg, Germany; mordasini@mpia.de
}

\affil{$^3$
University of Chinese Academy of Sciences, Beijing 100049, China
}

\affil{$^4$
Laboratoire Lagrange, UMR7293, Universit{\'e} de Nice Sophia-Antipolis, CNRS, Observatoire de la C{\^o}te d'Azur, 06300 Nice, France
}

\affil{$^5$
Key Laboratory of Planetary Sciences, Chinese Academy of Sciences, Nanjing 210008, China; jijh@pmo.ac.cn
}

\altaffiltext{}{
$^{\star}$Reimar-L\"ust Fellow of the MPG
}

\begin{abstract}

We apply hydrodynamic evaporation models to different synthetic planet populations that were obtained from a planet formation code based on a core-accretion paradigm. We investigated the evolution of the planet populations using several evaporation models, which are distinguished by the driving force of the escape flow (X-ray or EUV), the heating efficiency in energy-limited evaporation regimes, or both. Although the mass distribution of the planet populations is barely affected by evaporation, the radius distribution clearly shows a break at approximately 2 $R_{\oplus}$.  We find that evaporation can lead to a bimodal distribution of planetary sizes \citep{Owen2013} and to an ``evaporation valley" running diagonally downwards in the orbital distance - planetary radius plane, separating bare cores from low-mass planet that have kept some primordial H/He.  Furthermore, this bimodal distribution is related to the initial characteristics of the planetary populations because low-mass planetary cores can only accrete small primordial H/He envelopes and their envelope masses are proportional to their core masses.  We also find that the population-wide effect of evaporation is not sensitive to the heating efficiency of energy-limited description.  However, in two extreme cases, namely without evaporation or with a 100\% heating efficiency in an evaporation model, the final size distributions show significant differences; these two scenarios can be ruled out from the size distribution of $Kepler$ candidates.

\end{abstract}


\section{Introduction}

The large number of exoplanets detected thus far provides
significant observational constraints for theoretical studies on
planet formation and evolution. The population-wide distributions of
these exoplanets are closely related to their formation history. Based
on the core accretion paradigm
\citep{Lin1996,Weidenschilling1996,Ida2004,Mordasini2012a}, we can
generate synthetic planet populations that may explain the
statistical characteristics of exoplanets, such as the planet’s
semi-major axis versus its mass ($a$-$M$) \citep{Ida2004,Mordasini2009}
and its mass-radius diagram \citep{Mordasini2012b}.

Recently, the $Kepler$ mission detected thousands of
planetary candidates, the radii of which were measured using transit
observations \citep{Borucki2011}. These data may provide an important clue to the
size distribution of close-in planets
\citep{Howard2012,Dong2013,Petigura2013a}, which is an essential test
of the current planet formation theories. Previous planetary population
synthesis models by
\citet{Mordasini2012a,Mordasini2012b} reproduced a
synthetic planet population with a similar planet size distribution
as observed in the $Kepler$ data at radii $>$ 2 $R_{\oplus}$ .
However, for planets with radii $<$ 2 $R_{\oplus}$, a strong
decrease in the synthetic planet population is inconsistent with
the plateau in the planet size distribution of the
$Kepler$ data, after correction for
observational bias \citep{Howard2012,Dong2013,Petigura2013a}. The
divergence may be due to simplifications in the planetary
population synthesis models \citep{Mordasini2012a,Mordasini2012b}.
As a simplification, atmospheric evaporation after the protoplanetary disks have dissipated
\citep{Lammer2003,Baraffe2004,Murray-Clay2009},
which is important for close-in planets,
was not considered in the former studies.
Nevertheless, atmospheric evaporation may play a vital role in the
thermal evolution of low-mass planets;
this can alter the size distribution of close-in planets
with radii $<$ 2 $R_{\oplus}$ \citep{Owen2012,Lopez2012,Lammer2014}.
Thus in order to use the radius distribution of close-in
exoplanets to constrain planet formation models which is our final goal,
it is necessary to include the effect of atmospheric escape.

In the planetary evolution stage after disk
dissipation, the atmospheric structure and thermal evolution
of a close-in planet are strongly influenced by intense irradiation
from the host star
\citep{Guillot2002,Fortney2005,Fortney2008,Hansen2008,Guillot2010}.
Moreover,
the incident stellar X-ray and extreme-ultraviolet (EUV) flux 
can drive hydrodynamic evaporation in the atmosphere of close-in planets and 
yield a substantially higher mass-loss rate than 
for planets in the Jeans escape regime
\citep{Lammer2003,Baraffe2004,Murray-Clay2009,Owen2012,Lopez2012,Lammer2014}.
For example, oxygen, carbon, and magnesium were detected in the upper
atmosphere of HD 209458b at a distance of several planetary radii,
which indicates that its atmosphere is undergoing hydrodynamic escape
\citep{Vidal-Madjar2004,Vidal-Madjar2013}.
The estimated atmosphere
mass-loss rate for HD 209458b is $\gtrsim$ $10^{10}$\,g\,s$^{-1}$
\citep{Vidal-Madjar2004,Vidal-Madjar2008}. \citet{Wu2013} calculated
the masses of 22 sub-Jovian $Kepler$ planet pairs with the orbital
periods of $\sim$ 8 days using the TTV data and found that the mass-radius
relationship of these planets corresponds to the constant escape
velocity $\sim$20\,km\,s$^{-1}$, which is similar to the
sound speed of hydrogen plasma at $10^{4}$ K, indicating that
hydrodynamic evaporation is likely.

The atmospheric mass loss due to evaporation during planetary
evolution can be estimated by considering an energy-limited model.
This scenario assumes that a portion of the heating energy of stellar
irradiation contributes to $P$d$V$ work that expands
the upper atmosphere \citep{Watson1981}. The
mass-loss rate in the energy-limited model depends on the strength
of stellar XUV irradiation, including both X-ray and
EUV flux, and the heating efficiency, which
describes how much heating is converted to $P$d$V$
work. 
Following the temporal evolution of the XUV emission from a
sun-like star and assuming 100\% heating efficiency, early
studies showed that even a Jovian planet can lose
its entire initial envelope during its lifetime at close-in orbits
\citep{Lammer2003,Baraffe2004}.
However, later hydrodynamic simulations of
EUV-driven atmospheric escape show that although the EUV-driven mass
flow can produce the observed Lyman-$\alpha$ absorption signatures,
this mechanism can only evaporate a small portion of the mass of a
Jovian planet \citep{Tian2005,Yelle2004,Murray-Clay2009,Owen2012}.
Furthermore, \citet{Tian2005} showed that, in the energy-limited model,
the assumption that the incoming irradiation is absorbed inside a
single layer is inaccurate because the absorption depth of the
incident radiation depends on the wavelength.
\citet{Murray-Clay2009} found that
atmospheric evaporation is in a radiation-recombination-limited
regime for high EUV flux. In this regime, a large portion of the heating energy is
lost to cooling radiation, which decreases the mass-loss
rate. \citet{Owen2012} showed that atmospheric evaporation
can be either X-ray- or EUV-driven,
depending on the X-ray and EUV flux.
Most planets are in an X-ray-driven evaporation regime at the beginning
of evolution and then transition to an EUV-driven regime when the X-ray
flux falls below a critical value. This transition indicates that using the total XUV fluxes in an evaporation
model is insufficient because, in various regimes, the escape flow is dominated by
the stellar irradiation at different wavelengths, either X-ray or EUV \citep{Owen2012}.

Unlike Jovian planets, Neptune-like planets and super-Earths
are most likely to lose their
entire envelopes at small distances \citep{Lammer2009,Lopez2012,Owen2012}.
Because the
mass-loss timescale for planets scales with the planetary mass
$\times$ the planetary mean density, the planetary thermal
evolution coupled with atmospheric escape can elucidate the
observed threshold for low-density planets in the $M_{\rm p}^{2}/R_{\rm p}^{3}$ versus
distance$^{2}$ parameter space, above which there
is no low-mass, low-density transiting exoplanets have not been found \citep{Jackson2012,Lopez2012,Owen2013}.
Moreover, \citet{Owen2013}
found that evaporation can create a bimodal
planet size distribution around $2 R_{\oplus}$. In an extensive
investigation of evaporation that included thousands of
planets with different sizes and incident fluxes, \citet{Lopez2013}
recently showed that the bimodal distribution near $2 R_{\oplus}$ was
unclear.  In addition, they observed a diagonal region in the
semi-major axis versus radius ($a$-$R$) space,
where planets are relatively rare.

Herein, we simulate the thermal evolution of synthetic planet
populations that were obtained from a planet formation code
\citep{Alibert2005,Mordasini2012a,Mordasini2012b}. The atmospheric
mass-loss due to hydrodynamic evaporation is now included in the
planetary evolution. Thereby, we aim to determine how
evaporation statistically affects the entire planet population
and to make a more consistent comparison with the $Kepler$ data.
Our results show that the $a$-$R$ space is strongly influenced
by evaporation, which can modify the size distribution of the
planets within 1 AU.

The paper is organized as follows. In $\S$ \ref{sect:evapmodel}, we
describe the evaporation models used herein. Our
improved atmospheric model and numerical experiments on planet
evolution are shown in $\S$ \ref{sect:planetevolution}. In $\S$
\ref{sect:resultsII}, we present the mass and radius distributions of
the synthetic planet populations and compare the planet size
distribution of the synthetic populations with $Kepler$ candidates.
Finally, we provide a detailed discussion ($\S$ \ref{sect:discussion})
and brief summary ($\S$ \ref{sect:summary}).

\section{Evaporation models}

\label{sect:evapmodel}

The dominant heating source of hydrodynamic outflow
is either EUV or X-ray radiation,
which divides the hydrodynamic evaporation
into two distinct sub-regimes \citep{Owen2012}.
In this work, we use a model that includes both EUV-driven
and X-ray-driven evaporation through simple semi-analytical equations.

\subsection{X-ray-driven Evaporation}

X-ray photons have a smaller interaction cross-section compared with EUV photons;
Thus, they penetrate deeper into the planetary atmosphere.
In the early stage of planetary evolution, X-ray irradiation is
an important heating source for planetary atmospheres due to
strong X-ray emissions from young stars \citep{Ribas2005,Jackson2012}.
In this work, the mass-loss rate in an X-ray-driven regime is calculated using an
an energy-limited model and by assuming that part of the
heating energy is converted to $P$d$V$ work with
the efficiency factor $\epsilon$ \citep{Jackson2012}:
\begin{equation}
  \dot{m}=\epsilon\frac{\displaystyle{16 \pi F R^3_{\rm p}}}{\displaystyle{3 G M_{\rm p} K(\xi)}}
\label{Mdotxray}
\end{equation}
where $M_{\rm p}$ is the planetary mass,
$R_{\rm p}$ is the planet's radius at
the optical depth $\tau = 2/3$ in the thermal wavelengths.
($\tau$ is calculated using the grain-free Rosseland mean opacity
$\kappa_{\rm th}$ from \citet{Bell1994} and \citet{Freedman2008}),
$F$ is the X-ray flux in the wavelength range from
1 to 20 ${\rm \AA}$ from \citet{Ribas2005},
$\xi=R_{\rm roche}/{R_{\rm p}}$, and
\begin{equation}
K(\xi)=1-\frac{3}{2\xi}+\frac{1}{2\xi^3}
\end{equation}
accounts for the enhanced mass-loss rate by a factor of
$1/K(\xi)$ because the Roche lobe of a close-in planet can be
close to the planet's surface \citep{Erkaev2007}. 
\citet{Lammer2009} discuss the possible values of the efficiency factor
in energy-limited model and consider a realistic range of 0.1-0.25.
Considering that it is the X-ray from 5 to 10 ${\rm \AA}$ 
responsible for heating \citep{Owen2012}, the X-ray flux from 1 to 20  
${\rm \AA}$ used in our model might be too high.
Thus, we set the nominal efficiency factor in the X-ray-driven regime to 0.1.
We also set it to 0.2 for a comparison population synthesis 
to generate larger mass-loss rates.

\subsection{EUV-driven Evaporation}

EUV photons ($h\nu\ga 13.6$ eV) can ionize the hydrogen in
the upper atmosphere through the photoelectric effect.
Because the H recombination cooling peak lies slightly above $10^{4}$ K,
photoionization-recombination will produce an equilibrium
temperature of $\sim$ $10^{4}$ K in the ionized region \citep{Dalgarno1972}.
Based on the strength of the incoming EUV radiation, EUV-driven
evaporation can be further divided into two sub-regimes \citep{Murray-Clay2009}:

At high EUV fluxes with $F_{\rm EUV} > F_{\rm crit}$, 
a large fraction of the heating is lost to
cooling radiation.
The mass-loss rate increases slowly with the incoming flux as
$\dot{M} \propto F_{EUV}^{0.6}$ \citep{Murray-Clay2009}.
Therefore, the mass-loss rate cannot be estimated using a linear
energy-limited equation of incident flux.
Assuming that the escape flow is isothermal, the mass-loss rate is simply
the mass flux at the sonic point in the flow \citep{Murray-Clay2009}, we have
\begin{equation}
\dot{M}_{\rm rr-lim} \sim 4 \pi \rho_{\rm s} c_{\rm s} r_{\rm s}^2
\end{equation}
where $c_{\rm s} = [kT/(m_{\rm H}/2)]^{1/2}$ is the isothermal sound speed,
$T = 10^4$ K is the temperature at the sonic point,
$r_{\rm s} = GM_{\rm p}/(2c_{\rm s}^2)$ is the sonic point where
the mass flow escapes the planet at the sound speed, and $\rho_{\rm s}$ is
the flow density at the sonic point, which was
calculated as described in \citet{Murray-Clay2009}.

For $F_{\rm EUV} < F_{\rm crit}$,
the mass-loss rate is calculated using an energy-limited model that is
similar to Equation \ref{Mdotxray}.
The difference is that the $F$ used here is given by
the EUV fluxes of a sun-like star from \citet{Ribas2005}.
$R_{\rm p}$ in the EUV-driven regime is set as the planetary radius
where the optical depth becomes unity 
for EUV photons (20 eV),
at a pressure of $\sim$ 1 nanobar \citep{Murray-Clay2009}.
The efficiency factor is set by comparing Equation \ref{Mdotxray}
with Equation 19 in \citet{Murray-Clay2009},
which is an analytical equation used to estimate the mass-loss rates at
a low $F_{\rm EUV}$ and does not include the Roche lobe term $\xi$
\footnote[1]{The Roche lobe term is also neglected in the energy-limited EUV-
driven regime in our evaporation model.}
The efficiency factor is 0.06 in the nominal evaporation model and
0.12 in the comparison population synthesis.

We adopt the criterion from \citet{Murray-Clay2009},
which states that, for $F_{\rm EUV}$
$> 10^4~ {\rm\,erg}~ {\rm\,cm}^{-2}~ {\rm\,s}^{-1}$,
evaporation is in the radiation-recombination-limited regime for
hot Jupiters orbiting main-sequence solar analogs.
In reality, the critical $F_{\rm EUV}$ ($F_{\rm crit}$) for the transition from
a radiation-recombination- to an energy-limited evaporation
is a function of the planetary mass and radius, among other characteristics.
In our work, we simply choose a constant $F_{\rm crit}$
$= 10^4~ {\rm\,erg}~ {\rm\,cm}^{-2}~ {\rm\,s}^{-1}$ for each planet.
The variations of this critical value and its influence on 
the final planet population will be described in $\S$ \ref{modelruns}.

\subsection{Transition Between the X-ray- and EUV-driven Regimes}

Whether evaporation is X-ray- or EUV-driven depends on
the intensity of the EUV and X-ray irradiation received by a planet.
We use the criterion from \citet{Owen2012},
which separates X-ray- and EUV-driven regimes
based on whether the X-ray-driven flow can reach the sonic point
before it enters the ionization front.
This criterion can be determined based on the EUV luminosity of the host star.
The threshold EUV luminosity for the escape
flow in an EUV-driven regime is \citep{Owen2012}:
\begin{equation}
\begin{split}
  \Phi_* &\ge 10^{40} {\rm\, s}^{-1} \left(\frac{a}{0.1\rm { \,AU}}\right)^2\left(\frac{\dot{m}_{\rm X}}{10^{12}\rm { \,g\;\,  s}^{-1}}\right)^2\left(\frac{A}{1/3}\right) \\
&\quad \times \left(\frac{\beta}{1.5}\right)\left(\frac{R_{\rm p}}{10 R_{\oplus}}\right)^{-3}.
\end{split}
\end{equation}
where $\Phi_*$ is the EUV luminosity (in photons per second) of the host star,
$A$ (typically $\approx$ 1/3) is a geometric factor that approximates the steepness of
the density fall-off in the ionized portion of the
X-ray-heated flow \citep{Johnstone1998},
$\beta$ is the ratio of the X-ray sonic surface to the planetary radius
(it is of order unity \citep{Owen2012}; we set it to 1.5 herein),
and $\dot{m}_{\rm X}$ is the mass flux of the X-ray-driven flow that enters the ionization front.

\section{Planetary evolution}

\label{sect:planetevolution}

\subsection{The Synthetic Planet Populations}

The synthetic planet populations adopted herein are based on
a planet formation model
\citep{Alibert2005,Mordasini2012a,Mordasini2012b}, in which we
simulate the accretion of solid/gas materials and the disc-based
migration of a planet in combination with the evolution of a
protoplanetary gas disk. At 8 Myr, nearly all protoplanetary gas
disks have disappeared \citep{Mordasini2009}; thereafter, the
planets enter the evolution stage, where both gas accretion and
disk-driven migration end.
Planet-planet scattering and Kozai migration is not included in
the model because we use the one-embryo-per-disk approximation.
We take a
snapshot of each planet in a population at 10 Myr and set it as
the initial condition of the evolution stage. 
We use the following parameters
for each planet: core mass, envelope mass, luminosity,
mass-fraction of the ice in the core, initial deuterium fraction, and
semi-major axis.

The population-wide effect of atmospheric evaporation is our main
focus. We consider two variable quantities in population
synthesis: one is the description of the evaporation model that varies
with the dominant heating source (X-ray, EUV, or both) and
heating efficiency, and the other is the ISM
grain opacity reduction factor used during the formation stage. 
The grain opacity during the formation phase controls
the amount of H/He a core of given mass and luminosity
can accrete \citep{Movshovitz2010,Mordasini2014a}.
We follow the evolution of different synthetic planet populations
using different evaporation models. The details for each simulation
are listed in Table \ref{tab:simulist}. All populations are formed around
a 1 $M_{\odot}$ star and using the one-embryo-per-disk approximation.
A description of the concurrent formation of several planets can be found in \citet{Alibert2013}.
We define the nominal evaporation
model as one that includes both X-ray-and EUV-driven regimes.
The nominal synthetic population is defined as a population calculated
using the isothermal type I migration rate 
\citep{Tanaka2002} 
with a reduction factor
of 0.1 and a grain opacity reduction factor
($f_{\rm opa}$) of 0.003 compared with the full ISM value. The
low-mass, close-in planets in this population are formed based on a
protoplanetary disk that includes the stellar irradiation for
calculating the disk's temperature structure, and the combined
viscous and thermal criteria for the transition from type I to type
II migration. These assumptions are more realistic than the assumptions in
\citet{Mordasini2009}, wherein non-irradiated disks were used with only the
thermal criterion and a type I migration rate with a reduction
factor of 0.001. The inner boundary of the disk in our model is
0.1 AU due to limitations in the stellar irradiation description
for the disk \citep{Fouchet2012}. Radial velocity and transit surveys
have detected many planets within 0.1 AU, which demonstrates that
even close-in planets with orbital periods less than 10 days are
common \citep{Mayor2011,Dong2013,Petigura2013a}. Planets can reach
close-in orbits through disk-driven migration
\citep{Goldreich1980,Lin1996,Zhou2005} or migration due to
planetesimal disk dynamics
\citep{Terquem2007,Ji2011,Ormel2012}. A
close-in planet may also be formed through Kozai migration with tidal circularization
\citep{Wu2003,Fabrycky2007}, but this process can take considerably longer, and the
intense atmospheric evaporation stage may have passed. Because the inner boundary (0.1 AU) in the semi-major axis distribution of
our planet populations is too large to reasonably study 
hydrodynamic evaporation, we manually shift the semi-major axes
of all planets inward by 0.04 AU; thus, the planet populations begin
at 0.06 AU.

\begin{table}
\centering
\caption{Details of the different simulations}
\label{tab:simulist}
\begin{tabular}{lccc}
\hline
Simulation & Type I migration & $f_{opa}$  & Evaporation model \\
run    &  rate            &            &      \\
\hline
XE  & 0.1$\times$isothermal-rate  & 0.003    & X-ray + EUV \\
NoEV  & 0.1$\times$isothermal-rate  & 0.003    &  No Evap  \\
SatE  & 0.1$\times$isothermal-rate  & 0.003    &  EUV (Saturation) \\
XE2  & 0.1$\times$isothermal-rate  & 0.003    & XE $\times$ 2 \\
L12  & 0.1$\times$isothermal-rate  & 0.003    & \citet{Lopez2012} \\
B04  & 0.1$\times$isothermal-rate  & 0.003    & \citet{Baraffe2004} \\
\hline
NIOpa003  & non-isothermal   & 0.003    & X-ray + EUV \\
NIOpa0  & non-isothermal   & 0.0      & X-ray + EUV \\
NIOpa1  & non-isothermal   & 1.0      & X-ray + EUV \\
\hline
\vspace{0.05cm}
\end{tabular}
\end{table}

\subsection{Planet Structure Model}

A planet’s structure consists of a core and
a gaseous envelope. The planetary core mass is constant
during evolution. The core radius is determined by its mass,
its mass-fraction of ice, and the pressure at the bottom of the
gaseous envelope \citep{Mordasini2012b}. We assume spherical
symmetry in the planetary envelope and solve its structure by
combining the following one-dimensional hydrostatic equations:
\begin{subequations}
\begin{minipage}{.22\textwidth}
\begin{align}
\frac{{\rm d}m}{{\rm d}r}=4\pi r^{2}\rho \\
\frac{{\rm d}P}{{\rm d}r}=-\frac{Gm}{r^{2}}\rho
\end{align}
\end{minipage}
\hfill
\begin{minipage}{.22\textwidth}
\begin{align}
\frac{{\rm d}\tau}{{\rm d}r}=\kappa_{\rm th}\rho \\
\frac{{\rm d}L}{{\rm d}r}=0 
\end{align}
\end{minipage}
\end{subequations}
\\ where $r$ is the radius as measured from the planetary center,
$m$ is the cumulative mass inside $r$,
$\rho$ is the density in each spherical shell, $P$ the pressure,
$G$ the gravitational constant,
and $L$ is the planetary luminosity, which
includes radiogenic heating from the solid core.
We assume that the luminosity is constant with radius,
which does not significantly affect
the evolution as discussed in \citet{Mordasini2012a}.

The temperature gradient in the gaseous envelope depends on both
the optical depth and heat transfer mechanism
(convective or radiative) at each envelope layer.
We separate the gaseous envelope into two parts:
the atmosphere where most of the stellar irradiation is absorbed and
the envelope that lies below the atmosphere.
If an atmospheric layer is convectively stable,
we adopt the globally averaged temperature profile
from Equation (49) in the semi-grey model of \citet{Guillot2010} 
(which is derived using the Eddington approximation):
\begin{equation}
\begin{split}
T^4 &={3\tint^4\over 4}\left\{{2\over 3}+\tau\right\}+ {3\teq^4\over 4}\left\{{2\over 3}+\right. \\
    & \quad \left. {2\over 3\gamma}\left[1+\left({\gamma\tau\over 2}-1\right)e^{-\gamma\tau}\right]+ {2\gamma\over 3}\left(1-{\tau^2\over 2}\right)E_2(\gamma\tau)\right\}
\end{split}
\label{2bdglobal}
\end{equation}
where $\tint=(L/(4\pi\sigma_{B}R_{\rm p}^2))^{1/4}$ is the intrinsic temperature
that characterizes the heat flux from the planet's interior
($\sigma_{B}$ is the Stefan-Boltzmann constant),
$\teq=T_*(R_*/(2D))^{1/2}$ is the equilibrium temperature obtained by averaging
the stellar radiation over the entire planet surface ($T_*$ is the stellar temperature,
$R_*$ is the stellar radius, and $D$ is the distance from the planet to the star),
$\gamma=\kappa_{\rm v}/\kappa_{\rm th}$ is the ratio of the visible opacity to
the thermal opacity \citep{Guillot2010}.
The visible opacity $\kappa_{\rm v}$ is not explicitly calculated but
is incorporated in the model by $\gamma$.
$E_2(\gamma\tau)$ is the exponential integral $E_n(z)\equiv\int_1^\infty t^{-n}e^{-zt}dt$ with $n=2$.

The boundary between the atmosphere and envelope
should be at the optical depth in
visible wavelengths $\tau_{\rm v} \gg$ 1.
Based on the $\gamma$ defined in the semi-grey model,
we have $\tau \gg 1/(\sqrt{3}\gamma)$ at the transition \citep{Rogers2011}.
Because most of the starlight is absorbed at pressures less than 10 bar
\citep{Guillot2002}, we set the atmosphere/envelope boundary
at $\tau = 100/(\sqrt{3}\gamma)$ (which corresponds to a pressure of $\sim$ 10 bar).
If the envelope at $\tau > 100/(\sqrt{3}\gamma)$ is convectively stable,
the radiative temperature gradient is calculated using the diffusion
approximation that only includes the planet's intrinsic luminosity:
\begin{equation}
  \frac{{\rm d}T}{{\rm d}r}=-\frac{3\kappa_{\rm th}\rho L}{64\pi\sigma_{B}T^3 R^2}
\end{equation}

On the other hand, if a layer is convectively unstable
(i.e., the adiabatic temperature gradient is less steep than the radiative temperature gradient),
we use the adiabatic temperature profile instead:
\begin{equation}
  \frac{{\rm d}T}{{\rm d}r}=\frac{T}{P}\frac{{\rm d}P}{{\rm d}r}\left(\frac{{\rm ln}T}{{\rm ln}P}\right)_{\rm ad}
\end{equation}
where the adiabatic temperature gradient 
is calculated using the equation of state of \citet{Saumon1995}.
This convective adjustment is also used in the atmosphere, although
we do not allow detached convective zones there.

Planetary evolution is modeled using the framework of
\citet{Mordasini2012a,Mordasini2012b}, wherein 
the planetary luminosity $L$ and its temporal evolution is 
derived through energy conservation as $L = -{\rm d}E_{\rm tot}/{\rm d}t$,
where ${\rm d}E_{\rm tot}$ is the energy change due to gravitational contraction 
and release of internal heat (neglecting planetary rotation). 
The luminosity $L$ at each timestep controls the 
planetary structure and the changes of the interior adiabat,
and hence the temporal evolution of the planet.

\begin{table}
\centering
\caption{The $\gamma=\kappa_{\rm v}/\kappa_{\rm th}$ used in the atmospheric model.}
\begin{tabular}{cc}
\hline
Temperature (K) & $\gamma$ \\
\hline
260 & 0.005 \\
388 & 0.008 \\
584 & 0.027 \\
861 & 0.07 \\
1267 & 0.18 \\
1460 & 0.19 \\
1577 & 0.18 \\
1730 & 0.185 \\
1870 & 0.2 \\
2015 & 0.22 \\
2255 & 0.31 \\
2777 & 0.55 \\
\hline
\end{tabular}\\
\label{table:gamma}
\end{table}

\begin{figure}
\includegraphics[width=9cm]{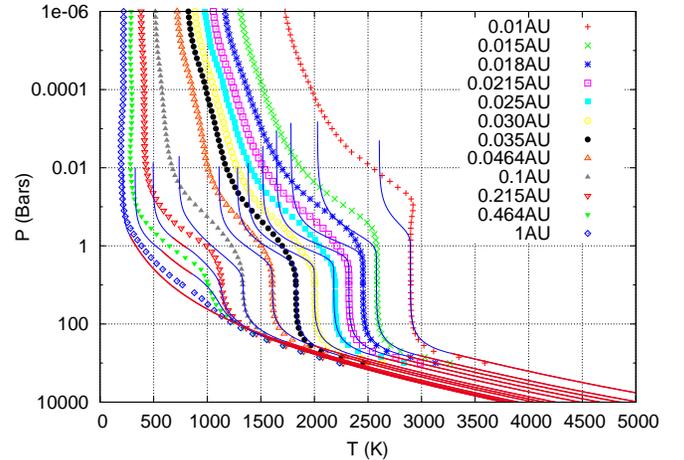}\\
\caption{Pressure-temperature models from \citet{Fortney2008} (points) and
the corresponding analytical fits (solid line) for different semi-major axes.
The modeled planets are with $g = 15$ m s$^{-2}$ and $T_{\rm int} = 200$ K,
and they orbit around a sun-like star.
The discrepancy in the upper part of atmosphere is due to the non-grey effects
\citep{Parmentier2014m,Parmentier2013s}.}
\label{fig:gamma}
\end{figure}

\begin{figure*}
\includegraphics[width=17.0cm]{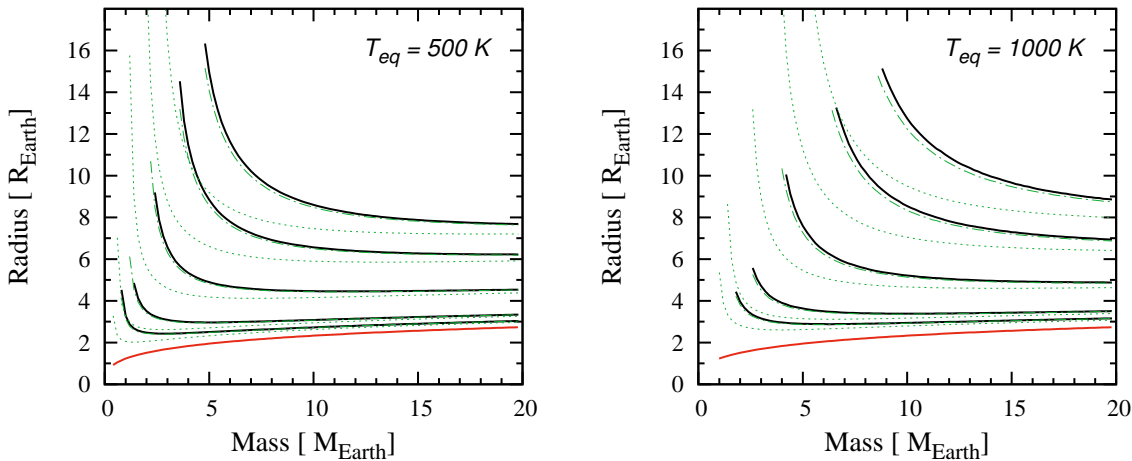}
\caption{
Mass-radius relationships for low-mass planets with
different envelope fractions $f_{\rm env/tot}$ ($M_{\rm env}/M_{\rm p}$).
The radius plotted here are the planetary radii at $\tau = 2/3$.
There are three groups of lines: The green dotted lines
are calculated using the fixed $\gamma = 0.6\sqrt{T_{\rm irr}/2000{\rm K}}$
and the planetary boundary at $\tau = 2/3$; 
these setting are the same as \citet{Rogers2011}.
The other two groups use the interpolated $\gamma$ at equilibrium temperatures at
500 and 1000 K, and the planetary boundary at $\tau = 0.01$
(the planetary radii plotted in the figure are at $\tau = 2/3$).
The green dot-dashed lines use the isotropic temperature profile (Equation 27 in \citet{Guillot2010}),
it is for comparison with the thick black lines that
use the globally averaged temperature profile (Equation 49 in \citet{Guillot2010}).
From bottom to top, the $f_{\rm env/tot}$ in each group is
0.001, 0.01, 0.1, 0.3, and 0.5, respectively.
The red line at the bottom is the mass-radius relationship of
the solid cores for planets with $f_{\rm env/tot}=0.001$.
In all of the runs, the luminosity of each planet is
set according to $L_{\rm p}/M_{\rm p} = 10^{-10.5}$ W\,kg$^{-1}$.
}
\label{fig:mrRg}
\end{figure*}

\subsection{Atmosphere Calibration}

\label{sect:resultsI}

One important parameter in the semi-grey model, $\gamma$,
determines how much of the incoming flux is absorbed in the upper
atmosphere. The $\gamma$ values used in our simulation are shown
in Table \ref{table:gamma} and are determined by comparing the
temperature in the deep isothermal zone of the analytical model
(Equation \ref{2bdglobal}) with the results from the line-by-line
radiative transfer models \citep{Fortney2005,Fortney2008,Parmentier2013s}. The
line-by-line EGP (Extrasolar Giant Planet) code was initially
developed by \citet{Mckay1989} to study Titan's atmosphere.
Since then, it has been extensively modified and adapted for
studies on giant planets \citep{Marley1999}, brown dwarfs
\citep{Marley1996,Marley2002,Burrows1997}, and hot Jupiters 
\citep[e.g.,][]{Fortney2005,Fortney2008,Showman2009}.

The $\gamma$ at a specific equilibrium temperature can be determined through
interpolation between these tabulated values. Figure \ref{fig:gamma}
shows the $PT$ profiles obtained by two different models.
The numerical profiles are calculated assuming 
a clear-sky, solar composition atmosphere where TiO and VO 
have rained out of the atmosphere \citep[see][]{Parmentier2013a,Parmentier2014m} and is neglected in the model. However, the effects of
a non-solar composition may be important but are not
considered in this study. 
The analytical solution is highly consistent with the line-by-line model in the deep atmosphere.
For pressures lower than 1 bar, 
the discrepancy is due to non-grey effects \citep{Parmentier2014m}.

\begin{figure*}
\includegraphics[width=17.0cm]{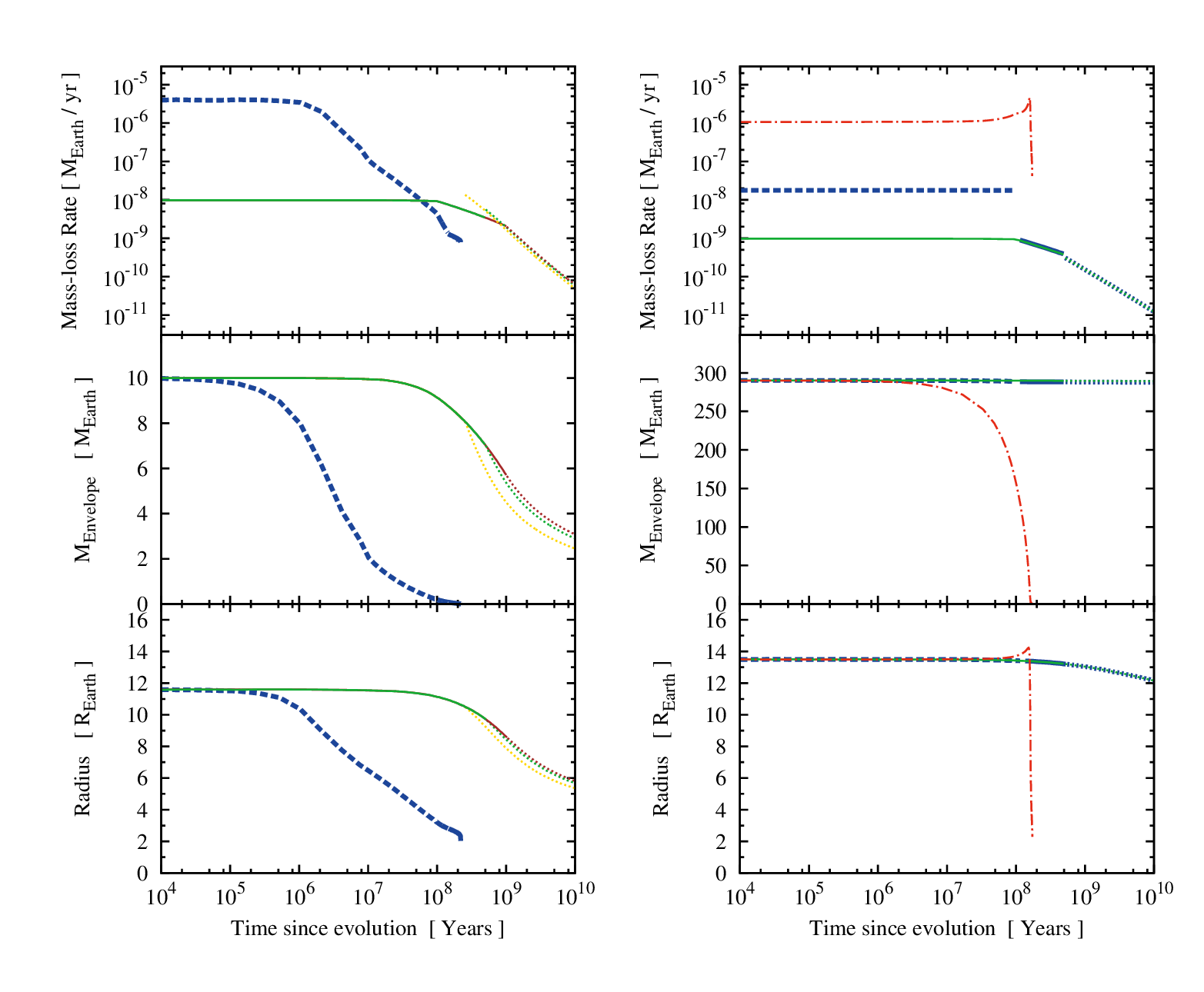}
\caption{
Temporal evolution of two close-in planets at 0.03 AU.
The left column shows the evolution of a Neptunian planet with
a 15 $M_{\oplus}$ core and a 10 $M_{\oplus}$ envelope.
The right column shows the evolution of a Jovian
planet with a 25 $M_{\oplus}$ core and a 290 $M_{\oplus}$ envelope.
The thick blue line is the experiment includes both
X-ray- and EUV-driven evaporation; the dashed part of the line indicates
the X-ray-driven regime, the solid part indicates
the radiation-recombination-limited EUV-driven regime,
and the dotted part indicates the energy-limited EUV-driven regime.
The thin green, yellow, and brown lines are the experiments 
that include only EUV-driven evaporation;
the solid part indicates the radiation-recombination-limited regime,
and the dotted part indicates the energy-limited regime.
The yellow and brown lines show the evolution of the
Neptunian planet using the $F_{\rm cirt}$
from a radiation-recombination-limited to an energy-limited regime
of $2 \times 10^4~ {\rm\,erg}~ {\rm\,cm}^{-2}~ {\rm\,s}^{-1}$
and $ 0.5 \times 10^4~ {\rm\,erg}~ {\rm\,cm}^{-2}~ {\rm\,s}^{-1}$.
The red dash-dotted lines in the right column use the
energy-limited model with a 100\% heating efficiency \citep{Baraffe2004},
}
\label{fig:single}
\end{figure*}

\subsection{Mass-radius Relationship of Low-Mass Planets}

We then compare the mass-radius relationships of the low-mass planets
from our model with those of \citet{Rogers2011}.
These low-mass planets have different envelope mass fractions
$f_{\rm env/tot}$ (the ratio of the envelope mass to the total planetary mass)
and are at an equilibrium temperature of 500 or 1000 K.
\citet{Rogers2011} use a fixed $\gamma$ of
$0.6\sqrt{T_{\rm irr}/2000{\rm K}}$ based on the fit in \citet{Guillot2010},
where $T_{\rm irr}=T_*(R_*/D)^{1/2}$ is the irradiation temperature.
\citet{Parmentier2014m} examine the fitted
$\gamma = 0.6\sqrt{T_{\rm irr}/2000{\rm K}}$ and find that it
is poor for temperatures lower than 1000 K due to
the disappearance of absorption by alkaline molecules (sodium, potassium, etc.).
\citet{Rogers2011} also set the planetary boundary at
an optical depth of $\tau = 2/3$.
As shown in Figure \ref{fig:mrRg}, the planetary radius is closely related
to the $\gamma$ value and the planetary boundary.
The mass-radius relationships calculated using the same $\gamma$ and
planetary boundary as \citet{Rogers2011} are highly consistent with their results.
We also show two other groups of mass-radius relationships that are
calculated using the tabulated $\gamma$ at 500 and 1000 K.
In addition to the different $\gamma$,
the planetary boundary for these two groups is set at
$\tau = 0.01$ (i.e., the atmospheric structure is integrated from $\tau = 0.01$;
notably, we adopt the radius at the optical depth $\tau = 2/3$
as the planetary radius, and for the figures herein,
the plotted planetary radii are at $\tau = 2/3$).
These two groups of mass-radius relationships are
considerably different from those of \citet{Rogers2011},
especially the low-mass planets
with a high $f_{\rm env/tot}$.
The difference between these two groups is the
temperature profile used in the semi-grey model
in the upper atmosphere.
One group uses the isotropically averaged temperature profile
in \citet{Rogers2011},
and the other group uses the globally averaged temperature profile, which
considers
the advective transport of energy to a certain extent \citep{Guillot2010}.
In contrast to the large discrepancies from the different
$\gamma$ and planetary boundary,
the differences between the mass-radius relationships
created by the isotropically averaged temperature profile and
the globally averaged profile are small.
The planetary radii calculated using the globally averaged temperature
profile are slightly larger than the radii calculated using the isotropic profile.
For the equilibrium temperature 500 K and with an $f_{\rm env/tot}$ of 0.1,
the radius of a 3 $M_{\oplus}$ planet calculated using the globally averaged
temperature profile is $\sim$ 4\% larger than the radius
for the isotropic profile,
and the difference between the radii from the two temperature profiles
is $<$ 1\% for planets larger than 6 $M_{\oplus}$.
The isotropic temperature profile is only used for
the comparison group shown in Figure \ref{fig:mrRg}.
In our population syntheses, planetary evolution is calculated
using the globally averaged temperature profile.

The red lines in Figure \ref{fig:mrRg} show the radii of the
solid cores of the planets with an $f_{\rm env/tot}$ of 0.001.
The large differences between the radii of the cores and the
total planetary radii show that the planetary envelope with only
0.1\% of the planetary mass produces a large increase in
the planetary radius, which is a well-known effect \citep[e.g.,][]{Adams2008}.
For example, at an equilibrium temperature of 500 K,
a planet with a total mass of approximately 1 $M_{\oplus}$ and
an envelope of 0.1\% of the planetary mass
will have a radius that is greater than 2 $R_{\oplus}$.
The planet's atmosphere is bloated due to its low surface gravity
and the heating from the incoming irradiation.
Consequently, the planetary radius will decrease dramatically after its
entire envelope is removed through evaporation.

\subsection{Illustrative Model Runs: Planetary Evolution with Escape}

\label{modelruns}

\begin{figure*}
\includegraphics[width=17.0cm]{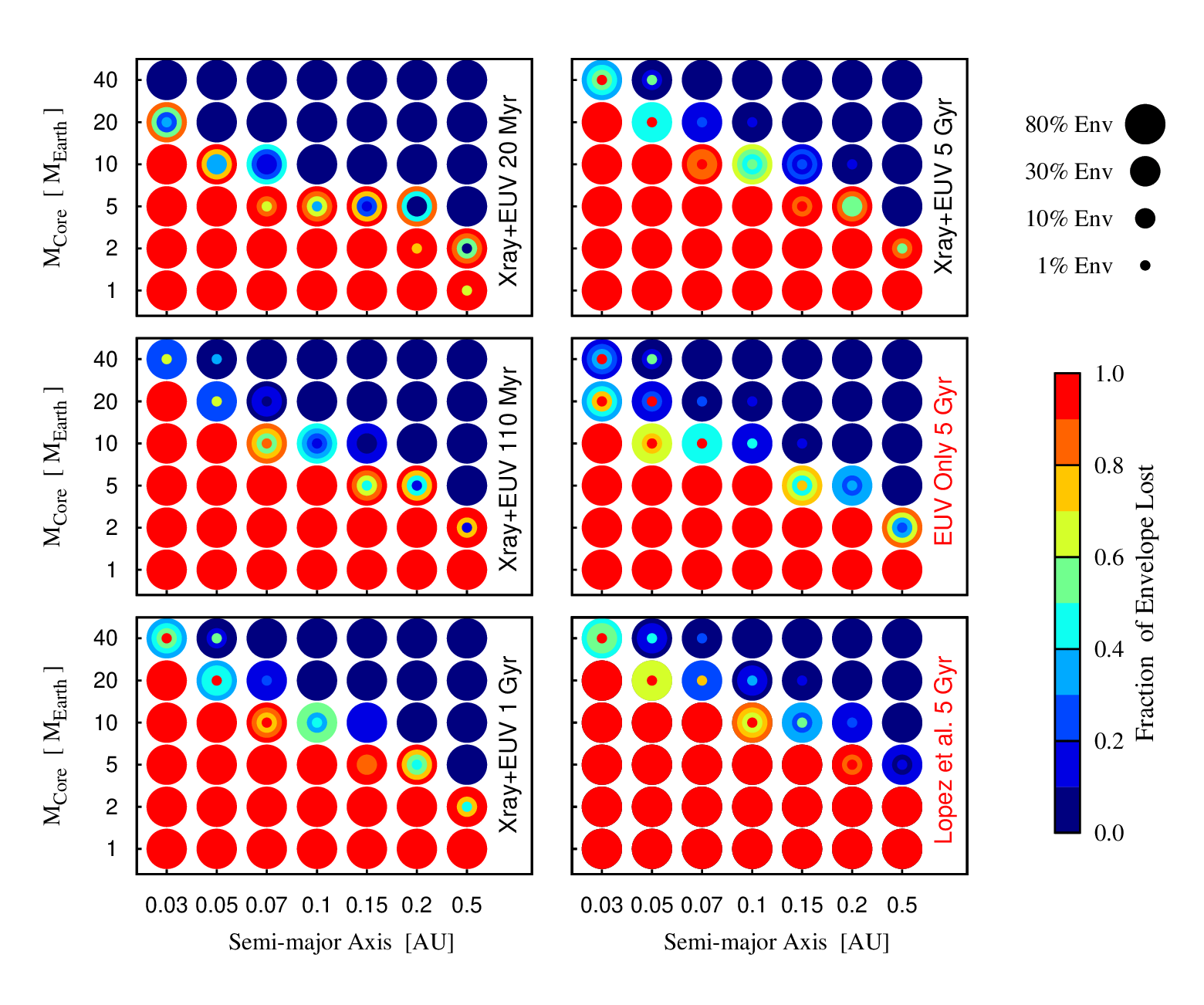}
\caption{
A parameter study of planetary evolution coupled 
with evaporation in the $M_{\rm core}$ versus
semi-major axis plane.
All planets orbit around a 1 $M_{\odot}$ star.
There are four planets (different point sizes) with 
different choices of $f_{\rm env/core}$ at each grid node, 
as indicated in the top right corner of the figure.
The color of each point shows how much of the initial envelope was lost.
The left column and the top panel in the right column show the
temporal evolution of the simulation using the nominal evaporation
model that includes both X-ray- and EUV-driven regimes.
The three panels in the right column compare the final results
of three simulations using different evaporation models:
both X-ray- and EUV-driven are included (nominal), EUV-driven only,
and the energy-limited model from \citet{Lopez2012}.
}
\label{fig:grid}
\end{figure*}

Figure \ref{fig:single} shows the evolution of
a Neptunian planet and a Jovian planet located at 0.03 AU.
The atmospheric escape is included such that
the planetary envelope mass is reduced at each timestep
based on the calculated mass-loss rate.

The Neptunian planet, with a 15 $M_{\oplus}$ core and 10 $M_{\oplus}$ envelope,
loses its entire initial envelope in the first 220 Myr when both
X-ray- and EUV-driven evaporation are included.
If only EUV-driven evaporation is included,
the planet can retain an envelope of $\sim$ 2.9 $M_{\oplus}$
after 10 Gyr of evolution.
X-ray-driven evaporation in the early stage of
planetary evolution is efficient at removing gas from a planet
because the planetary atmosphere is bloated and the
X-ray flux from the host star is intense \citep{Ribas2005}.
The mass-loss rates in the first 1 Myr of planetary evolution
are approximately $4 \times 10^{-6} M_{\oplus}/yr$.
From $\sim 1.45 \times 10^{8}$ years and forward,
evaporation transitions to the EUV-dominated
radiation-recombination-limited regime.
At this time, the planet only has a 0.085 $M_{\oplus}$ envelope remaining.
Soon thereafter, the planet loses its entire envelope and becomes a
15 $M_{\oplus}$ rocky core with a $\sim$ $2 R_{\oplus}$ radius.
In contrast, if only EUV-driven evaporation is included, in its early stage,
the planet is in the radiation-recombination-limited regime wherein most
incoming energy is lost to cooling radiation.
The mass-loss rates in this regime are determined based on the density of
the escape flow at the sonic point; they are below
$10^{-8} M_{\oplus}/yr$ in the first 100 Myr of planetary evolution.
The transition from radiation-recombination-limited evaporation to
energy-limited evaporation occurs at approximately $5.1 \times 10^{8}$ years and
when the EUV flux from the star is below
$10^4~ {\rm\,erg}~ {\rm\,cm}^{-2}~ {\rm\,s}^{-1}$.
After 10 Gyr, the final bulk composition of this planet includes
a 15 $M_{\oplus}$ core with a 2.9 $M_{\oplus}$ envelope,
and the planetary radius is 5.67 $R_{\oplus}$.
The discontinuous change in the mass-loss rate for this experiment demonstrates
that the critical EUV flux, which delimits the radiation-recombination-limited
and energy-limited evaporation regimes, is in principle inappropriate for this Neptunian planet.
In Figure \ref{fig:single},
we also show the evolution of the same planet but using different
$F_{\rm crit}$ at $2 \times 10^4$ and
$ 0.5 \times 10^4 ~ {\rm\,erg}~ {\rm\,cm}^{-2}~ {\rm\,s}^{-1}$.
These two experiments generate a final planetary radius of 5.34 $R_{\oplus}$
($F_{\rm crit} =
2 \times 10^4 ~ {\rm\,erg}~ {\rm\,cm}^{-2}~ {\rm\,s}^{-1}$)
and 5.81 $R_{\oplus}$ ($F_{\rm crit} =
0.5 \times 10^4 ~ {\rm\,erg}~ {\rm\,cm}^{-2}~ {\rm\,s}^{-1}$),
which corresponds to the 5.8\% and 2.5\% changes in the radius, respectively.
Notably, these two experiments only include EUV-driven evaporation.
When X-ray-driven evaporation is included in the model,
most of the planets that could be evaporated to bare cores
lost their entire envelope when the evaporation remained
in the X-ray-driven regime.
Thus, the constant $F_{\rm crit}$ used for
the low-mass planets in our model will not excessively affect the
population-wide radius distribution, as shown in the following section.

For the Jovian planet, the evaporation models with or without X-ray-driven
mass-loss do not show significantly different final results.
The modeled planet has a 25 $M_{\oplus}$ core and a 290 $M_{\oplus}$ envelope.
As shown in the right column of Figure \ref{fig:single}, the evolution
of this planet under different evaporation models overlaps
during the 10 Gyr's evolution.
The only notable difference between the two runs is that, in the first 100 Myr,
the mass-loss rate in the X-ray-driven regime is approximately 18-fold greater than
in the EUV-driven, radiation-recombination-limited regime.
This difference is insufficient for producing a notable change in the planetary
mass and radius because the total mass of the envelope lost
only composes a small fraction of the planetary mass.
In the end, the experiment that also included X-ray-driven evaporation
lost $\sim$ 2.29 $M_{\oplus}$ of the envelope, and the other run, which did not
include X-ray-driven evaporation, lost $\sim$ 0.72 $M_{\oplus}$ of the envelope.
The only way for a Jovian planet to lose a significant portion of
its envelope is by assuming an energy-limited evaporation model
with a 100\% heating efficiency \citep{Baraffe2004}.
In this case, the Jovian planet began with a mass-loss rate greater than
$10^{-6} M_{\oplus}$\,yr$^{-1}$. 
With this high mass-loss rate,
the planetary radius expands at approximately $10^8$ years due to the
entropy change in the outer radiative layers.
Detailed descriptions of this interesting process can be found in \citet{Baraffe2004,Hjellming1987}.
In turn, the larger planetary radius results in an even greater mass-loss rate.
Beginning at $\sim 1.5\times10^{8}$ years, the planet reaches
a runaway mass-loss stage, as observed by \citet{Baraffe2004}.
Eventually, the planet loses its entire initial envelope after
$\sim 1.72\times10^{8}$ years of evolution.

\subsection{A Parameter Study}

In Figure \ref{fig:grid}, we show a parameter study on evaporation.
We consider three parameters: the planetary semi-major axis,
the planetary core mass,
and the ratio of the planetary envelope mass to the core mass
($f_{\rm env/core}$).
The parameter space is similar to \citet{Lopez2013}, where
the incident flux from the star is used instead of the planetary semi-major axis
because the distance from a planet to its host star essentially determines
the incident flux that the planet can receive for a fixed stellar type.
The modeled planets are located at 0.03-0.5 AU with a core of 1-40 $M_{\oplus}$.
At each semi-major axis and each core mass, four planets have
different $f_{\rm env/core}$, i.e., 1\%, 10\%, 20\%, and 80\%.
We do not consider planets with a larger $f_{\rm env/core}$.
Due to the long Kelvin-Helmholtz timescale,
it is unlikely that a small core will accrete
significant levels of gas during the formation stage.
A massive core may accrete high levels of gas,
but the effect of evaporation on gas giants is small,
as shown in Figure \ref{fig:single}; thus,
gas giants are not included in this parameter study.
Because all planets in this parameter study are artificial bodies without
an initial luminosity from the
self-consistent planet formation stage, we set their
initial luminosities to the value that corresponds to an entropy
at the core-envelope boundary that equals
$7.11\times(M_{\rm core}/M_{\oplus})^{0.0422}\times (f_{\rm env/core})^{0.0175}$.
This is an empirical fit of the central entropies of the planets
with a 1-40 $M_{\oplus}$ core and a $f_{\rm env/core}$ of 1\%-80\%
in our synthetic population at 10 Myr.
This fit shows that the initial entropy of a planet 
increases with the core mass and envelope to core mass fraction. 
The initial entropy of low-mass planets is an interesting subject
that will be investigated separately in future work.

\begin{figure*}
\includegraphics[width=17.0cm]{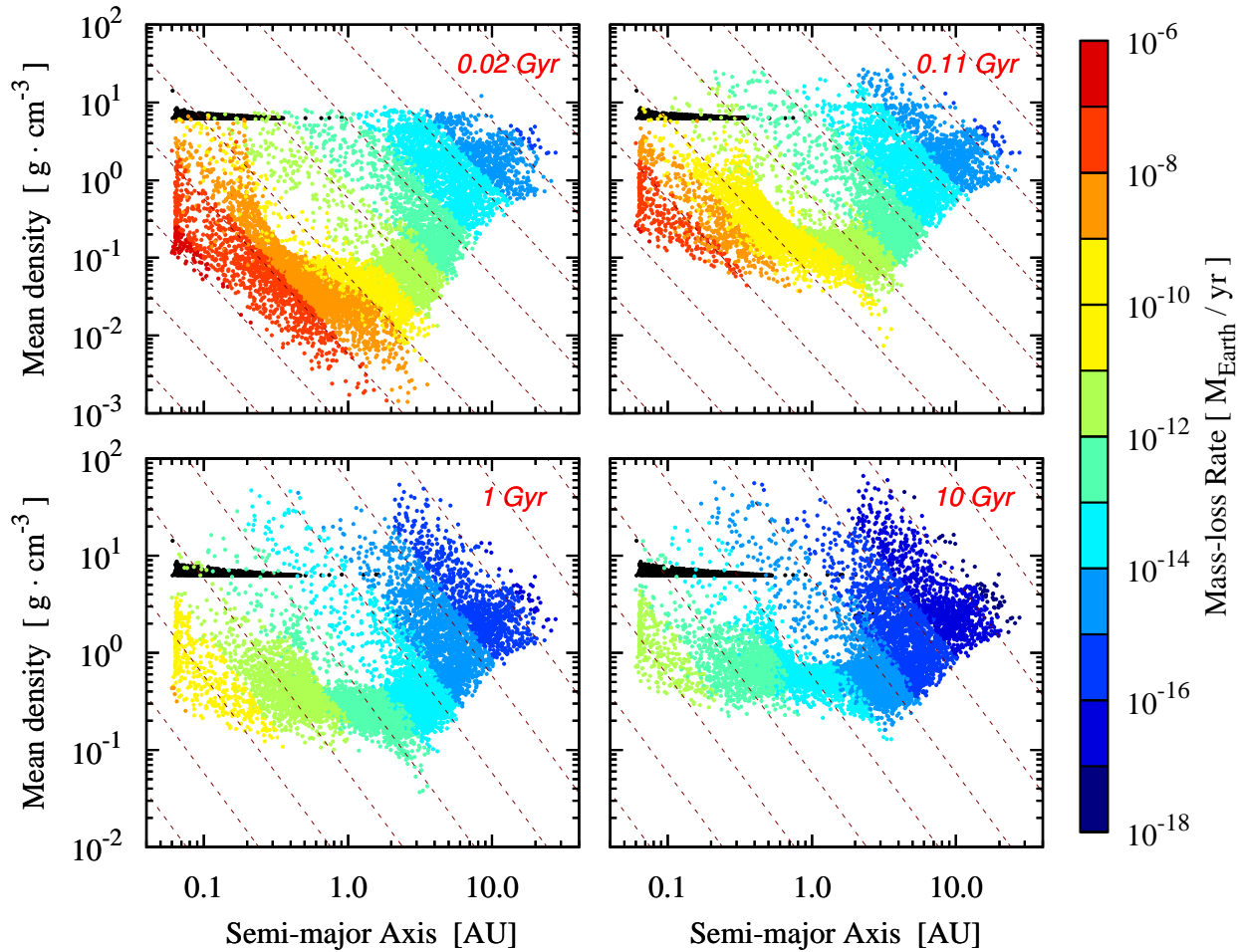}
\caption{
Temporal evolution of the mass-loss rates of the reference simulation
in the planetary semi-major axis versus mean density plane.
All planets orbit around a 1 $M_{\odot}$ star.
Each point in the figure corresponds to a planet in the synthetic population.
The color of each point show the mass-loss rate of the planet.
The black points are the planets that have lost all their initial envelopes.
The parallel dashed lines show the loci of identical
mass-loss rates in the energy-limited evaporation regime,
i.e., points along each line will have the same mass-loss rate.
}
\label{fig:arhoMdotfid}
\end{figure*}

\begin{figure*}
\includegraphics[width=17.0cm]{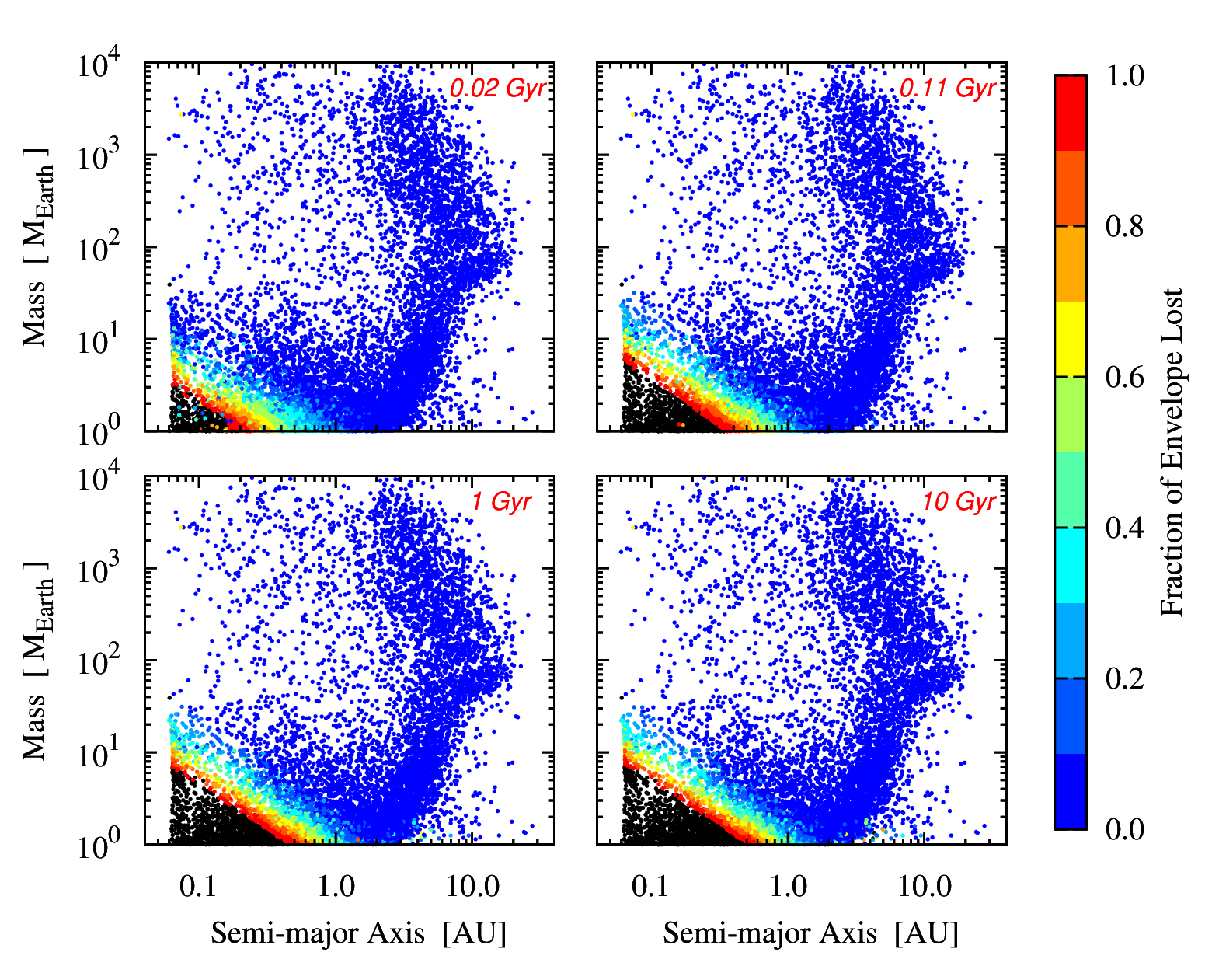}
\caption{
The temporal evolution of the planetary mass and semi-major axis distribution
of the reference simulation.
The color of each point shows how much of the initial envelope was lost
($M_{\rm lost}/M_{\rm initial}$).
The black points are the planets that have lost all their initial envelopes.
Note the large population of close-in, low-mass planets that
have evaporated to bare cores.
}
\label{fig:aMfid}
\end{figure*}

The left column of Figure \ref{fig:grid} shows the evolution of
the simulation using the nominal evaporation model, which includes
both the X-ray- and EUV-driven mechanisms.
Four snapshots of this simulation (at 20 Myr, 110 Myr, 1 Gyr, and 5 Gyr)
are presented.
At 20 Myr (after 10 Myr of evolution), all of the planets in the bottom left corner have lost their entire envelopes and have become bare rocky cores.
There are two reasons for this observation:
one, the large planetary radii due to the heating effect from the
intense stellar irradiation and large planetary intrinsic entropies
during the early stage and,
two, the manually fixed $f_{\rm env/core}$ for small cores
can be substantially higher than predicted in the formation calculations.
In particular, the formation calculations do not produce
cores of 1, 2, or 5 $M_{\oplus}$ with
a $f_{\rm env/core}$ of 80\% at 0.03 or 0.05 AU.
The high temperature at the outer boundary and
large initial luminosities for these small planets with a large
$f_{\rm env/core}$ appear to produce an unstable envelope structure
that cannot be modeled using the hydrostatic equilibrium approximation.
The only structure that we find for these planets is a bloated
atmosphere that expands beyond the Hill sphere, which is unstable.
Thus, their initial atmospheres evaporate in a short time.
\citet{Kurokawa2014} show comparable cases with low-mass planets undergoing a similar dynamic, i.e.,
runaway mass-loss.
After the early, intense, X-ray-driven evaporation stage,
the evaporation transitions to the moderate radiation-recombination-limited
or energy-limited EUV-driven regimes and the snapshots at the 1 Gyr and 5 Gyr only slightly differ.

The three panels in the right column compare the final status
of the three simulations, with
the only difference between them being the evaporation model.
One uses the nominal model, which includes both X-ray- and EUV-driven mechanisms,
another only uses the EUV-driven evaporation model, and a third
uses the same energy-limited evaporation model as that in \citet{Lopez2012}.
The final configurations of these three groups are similar:
the planets in the bottom left corner become bare rocky cores,
while planets with large cores at large semi-major axes
retain most of their initial envelopes.
The differences between these three groups lie in the diagonal region
of each panel.
X-ray-driven evaporation in the early stage is more
effective at removing planetary atmosphere than EUV-driven evaporation;
thus, the diagonal region in the
simulation using the nominal evaporation model includes more bare cores than
the simulation using only the EUV-driven mechanism.
The simulation
using the energy-limited evaporation model in \citet{Lopez2012} produces the greatest mass-loss.
Consequently, it includes the fewest planets with portions of their initial envelopes, and the $f_{\rm env/core}$
of each planet at 5 Gyr is the smallest of the three simulations.

\begin{figure*}
\includegraphics[width=17.0cm]{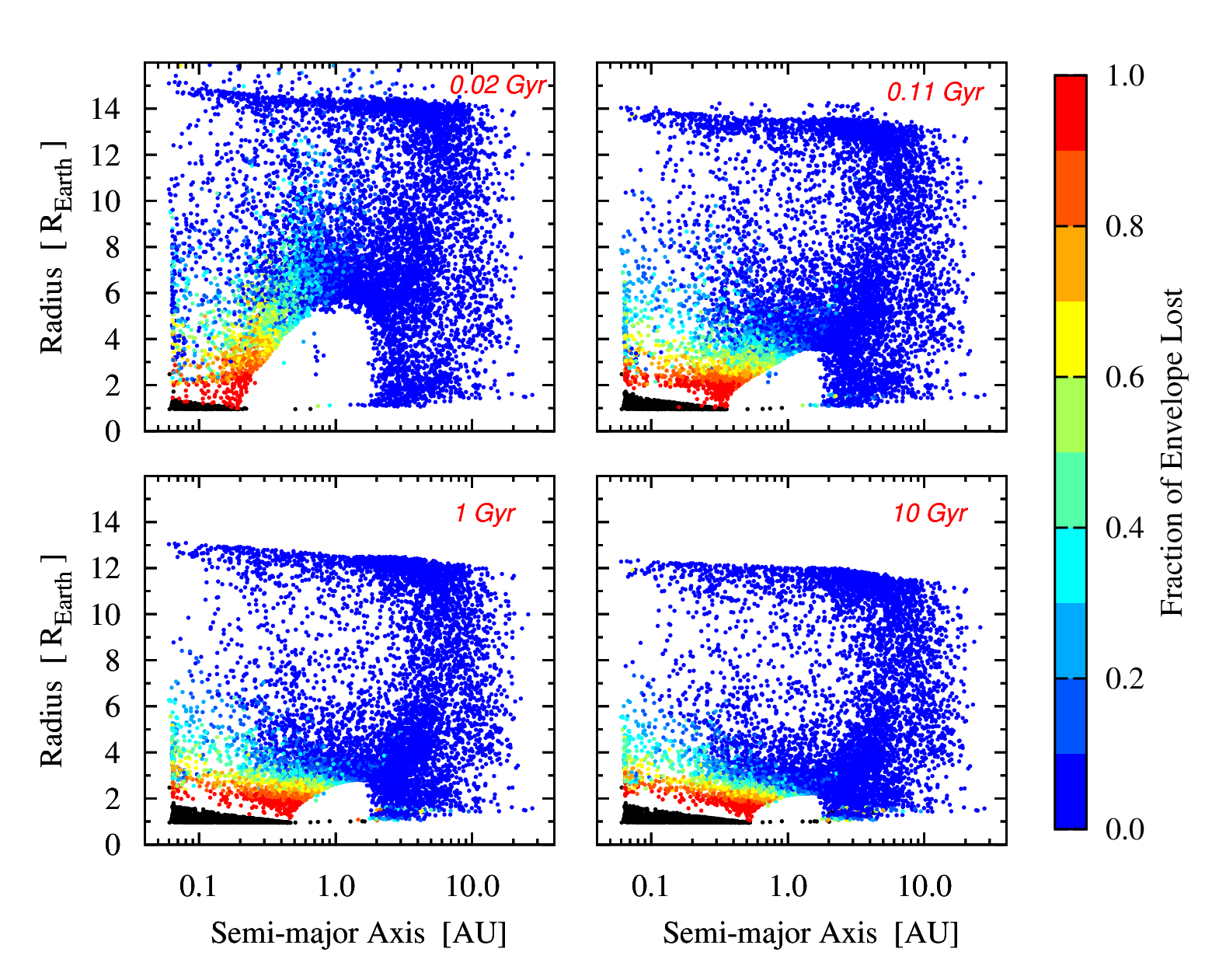}
\caption{
Temporal evolution of the planetary radius and semi-major axis
distribution of the reference simulation.
The color of each point shows how much of the initial envelope was lost.
The black points are the planets that have lost all their initial envelopes.
Above the black points, a separating ``evaporation valley" that runs
diagonally downward from 0.06 to 0.5 AU at 5 Gyr is clearly visible.
The empty region at intermediate orbital distances (extending at 0.02 Gyr from 0.2 to 2 AU)
is in contrast an artifact of assuming a minimal planetary mass of 1 $M_{\oplus}$,
and has no physical meaning.
Note that all planets start with a primordial H/He envelope.
In reality, this is likely not the case for all low-mass planets.
}
\label{fig:aRfid}
\end{figure*}

\section{The population-wide impact of evaporation}

\label{sect:resultsII}

\begin{figure*}
\includegraphics[width=18.0cm]{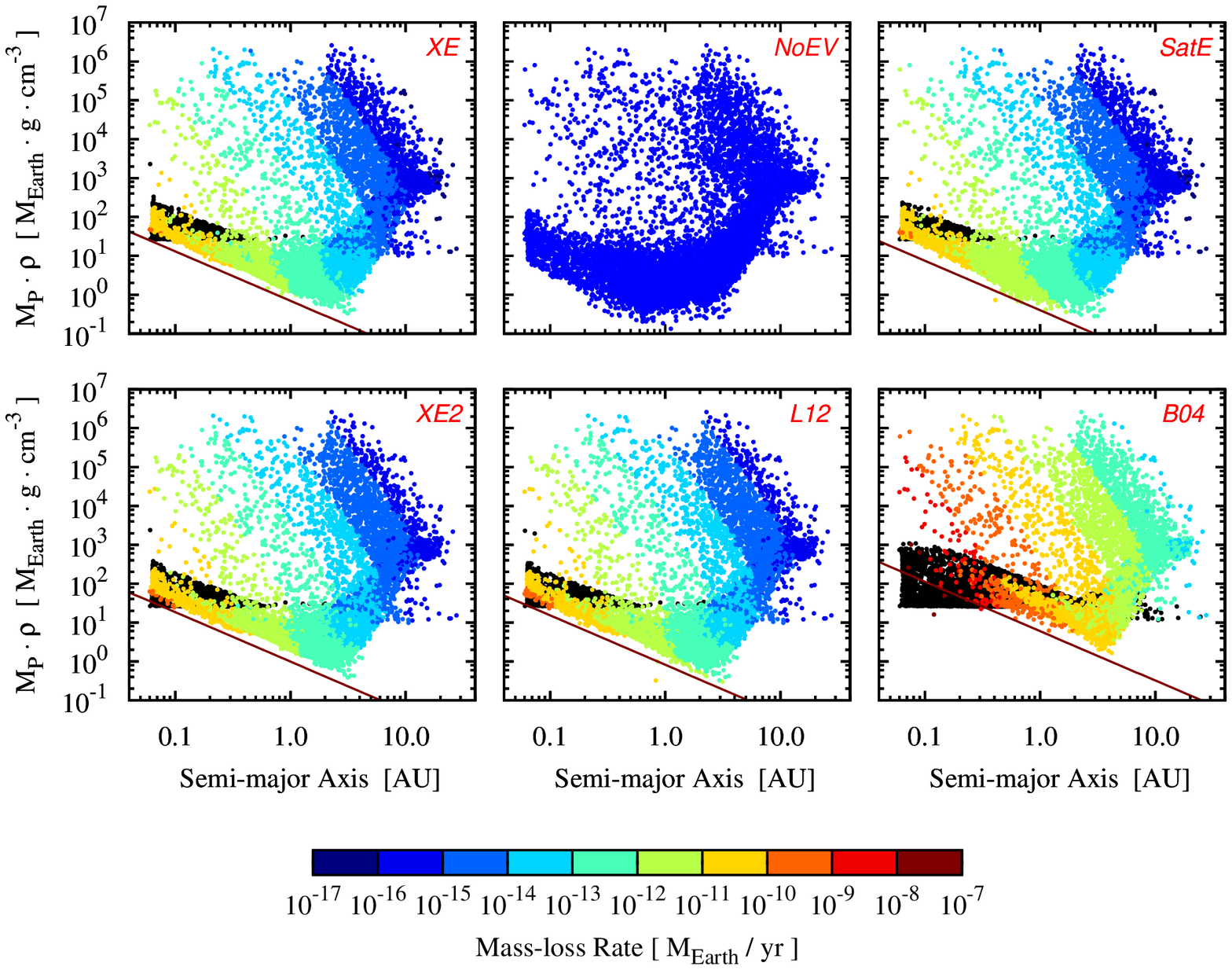}
\caption{
The mass-loss rates of the nominal planet population using
different evaporation models (Table \ref{tab:simulist}) at 5 Gyr
in the $M_{\rm p}\cdot\bar{\rho}$ versus semi-major axis plane.
The black points are the planets that have lost all their initial envelopes.
The minimal planetary mass is 1 $M_{\oplus}$.
The solid line in each panel indicates the evaporation threshold.
}
\label{fig:6arhoMdot}
\end{figure*}

\begin{figure*}
\includegraphics[width=18.0cm]{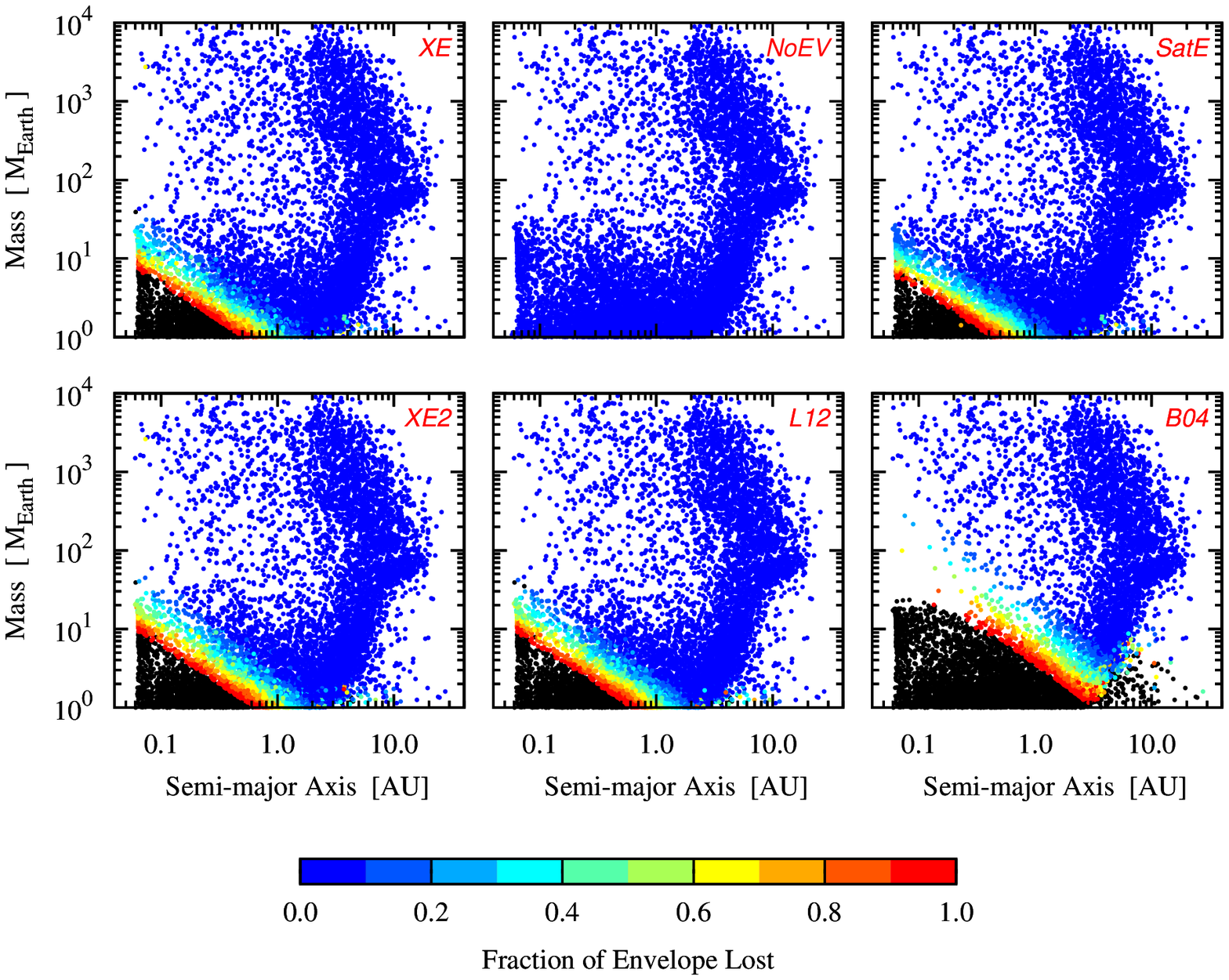}
\caption{
The planetary mass versus semi-major axis distributions of 
the nominal planet population
using different evaporation models at 5 Gyr (Table \ref{tab:simulist}).
The color of each point shows how much of the
initial envelope was lost.
The black points are the planets that have lost all their initial envelopes.
}
\label{fig:6aM}
\end{figure*}

\begin{figure*}
\includegraphics[width=18.0cm]{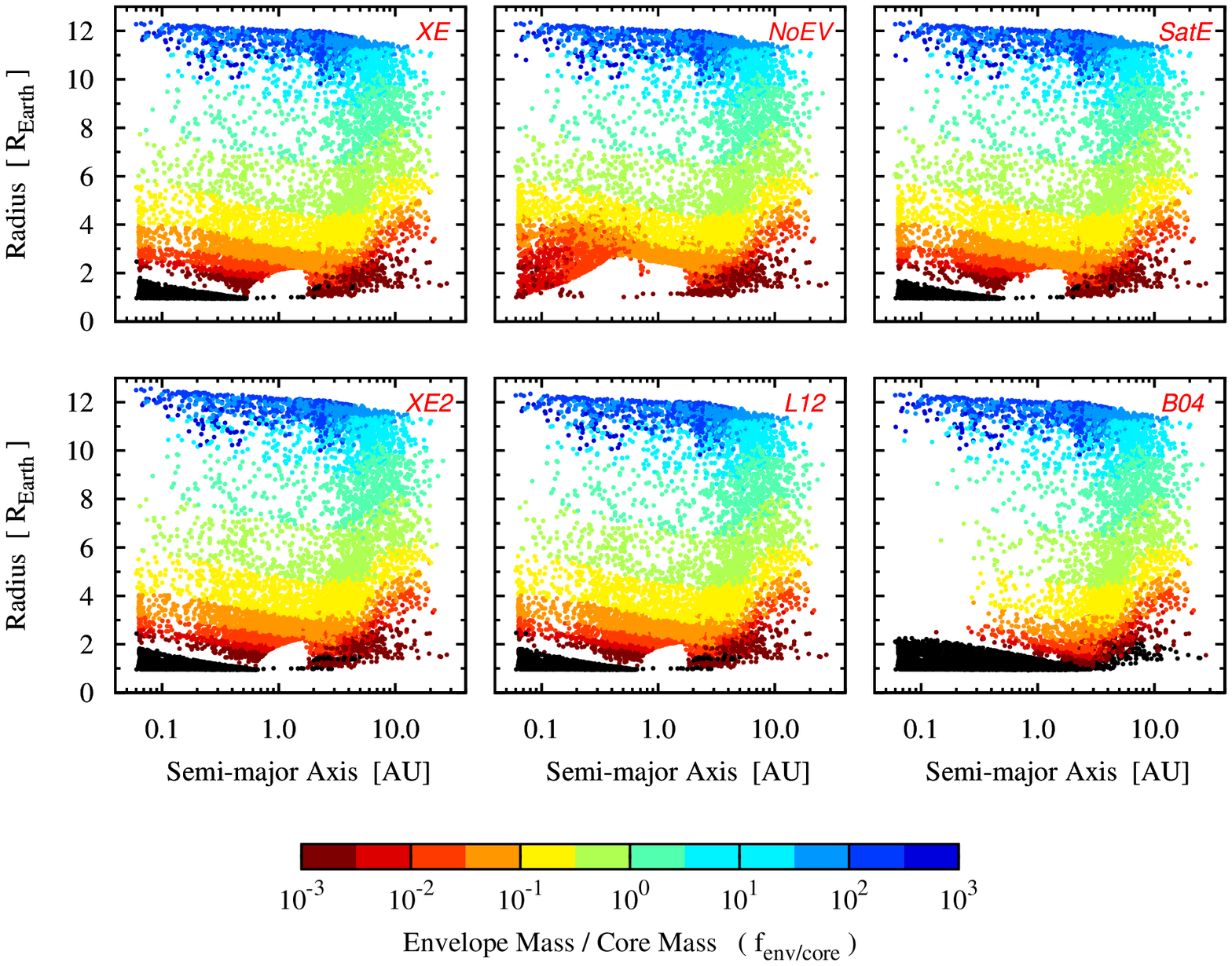}
\caption{
The planetary radius versus semi-major axis distributions of 
the nominal planet population
using different evaporation models at 10 Gyr (Table \ref{tab:simulist}).
The color of each point shows the ratio of the planetary envelope mass
to the core mass, $f_{\rm env/core}$.
The black points are the planets that have lost all their initial envelopes.
It is obvious that the simulation B04 is incompatible with the observed $a$-$R$ relationship.
}
\label{fig:6aR}
\end{figure*}

We then couple planetary population synthesis with
different evaporation models to determine the population-wide
effect of atmospheric evaporation.
First, in Figure \ref{fig:arhoMdotfid}, \ref{fig:aMfid},
and \ref{fig:aRfid}, we show the evolution of the mass-loss rates,
mass and radius distribution of our nominal planetary population
using the nominal evaporation model.
We then investigate the influence of the efficiency of evaporation mechanism.
In Figure \ref{fig:6arhoMdot}, \ref{fig:6aM},  \ref{fig:6aR}, and \ref{fig:6mr}, 
we compare the mass-loss rates, mass and radius distributions,
and mass-radius relationships using
different evaporation models (or without evaporation).
The effect of the different grain opacities used during the
planetary formation stage is demonstrated by the mass-radius relationships of
three different planet populations in Figure \ref{fig:nio}.
Finally, we compare the radius distribution of our synthetic populations
with the $Kepler$ data in Figure \ref{fig:3histo} and \ref{fig:9histo}.

\subsection{Synthetic Planets: A, the Reference Simulation}

\label{sect:nominal}

In Simulation XE (see Table \ref{tab:simulist}),
we apply the nominal evaporation model
(X-ray + EUV) to the nominal planetary population;
therefore, it is referred to as our reference simulation.
In Figure \ref{fig:arhoMdotfid}, we plot the temporal evolution of
the planet mass-loss rates for Simulation XE
in the planet's semi-major axis versus mean density plane.
The evaporation rates of close-in planets are large at 0.02 Gyr and 0.11 Gyr.
At 0.02 Gyr, close-in planets at 0.06 AU can have a mass-loss rate of
$\sim$ $10^{-6}$ $M_{\oplus}$ yr$^{-1}$.
The mass-loss rates for most of planets beyond 1 AU are
less than $10^{-10}$ $M_{\oplus}$ yr$^{-1}$.
At 1 Gyr, nearly all planets that retain portions of their
envelope have a mass-loss rate of less than $10^{-10}$ $M_{\oplus}$ yr$^{-1}$,
which explains why the envelope mass fractions and radius distributions
in Figure \ref{fig:aMfid} and Figure \ref{fig:aRfid} (in the following section)
barely change after 1 Gyr.
The parallel dashed lines show identical mass-loss rates
for a purely energy-limited evaporation model (i.e., the planets along each line
have the same mass-loss rate as those in the energy-limited regime); the
mass-loss rate is a function of the planet’s mean density and incoming flux.
We use an energy-limited model in both the X-ray and the low EUV regimes;
thus, most of the color belts in Figure \ref{fig:arhoMdotfid} are parallel to these dashed lines.
At 0.02 Gyr, a considerable portion of the planets are
in the radiation-recombination-limited regime at $\sim$ 0.2 to 1 AU;
hence, the yellow and orange belts slightly deviate
from the parallel dashed lines.
From 1 Gyr, nearly all of the planets are in the energy-limited,
EUV-driven regime, and the color belts closely follow the parallel dashed lines.

Figure \ref{fig:aMfid} shows the temporal evolution of the mass versus
the semi-major axis ($a$-$M$) distribution of the reference simulation
(the XE simulation). The $a$-$M$ distribution shows typical
sub-populations, such as the numerous low-mass ``failed cores" that only
accrete a limited quantity of gas, as well as many giant
planets that preferentially form outside of the snow line
\citep{Ida2004,Mordasini2009}. The color of each point in Figure
\ref{fig:aMfid} denotes the fraction of the initial envelope that
is lost for this planet (i.e., 1.0 indicates that a planet
has lost all of its initial H/He, and 0 means that the planet retains
its entire envelope). The black region in the bottom right corner corresponds to
the planets that have lost their entire initial envelope. The snapshot
at 0.11 Gyr presents a greatly increased black bare-core region;
however, the further increase is small for 1 and 5 Gyr.
The critical planetary mass below which a planet
can lose its entire envelope ($M_{\rm crit}$) has the form $M_{\rm crit}(t)=M_{\rm
crit}(t,a_{0}=0.06 ~{\rm AU})(a/a_0)^{-1}$. For
example, at t = 10 Gyr the critical planetary mass is $M_{\rm crit}(a=0.06 ~{\rm AU})\simeq 10 M_{\oplus}$, while
$M_{\rm crit}(a=0.7 ~{\rm AU})\simeq 1 M_{\oplus}$. Most of the
fully depleted planets are low-mass ``failed cores" ($M \leq 10
M_{\oplus}$) that only accrete a small envelope during the
formation stage. Although these planets have evaporated to bare, rocky
cores, they do not lose a substantial amount of their total mass because their initial envelope masses are
substantially lower than their core masses. Neptunian planets have
greater initial envelope mass fractions, but it is difficult for them to
lose a large portion of their envelopes. All of the Jovian planets retain
most of their envelopes. Thus, the
$a$-$M$ distribution of the entire planet population exhibits nearly no change: the four
snapshots in Figure \ref{fig:aMfid} all have similar
shapes. Notably, herein we only include planets with $a \geq 0.06 {\rm
AU}$; these results will differ for planets at very close-in orbits.

However, the $a$-$R$ distribution
of the entire planetary population is clearly modified by evaporation.
Figure \ref{fig:aRfid} shows the $a$-$R$ distribution of the same
reference simulation.
First, certain features are related to the
planet formation and evolution model.
For example, the empty region from 0.2 to 2 AU at 0.11 Gyr
is an artifact of using a minimal core mass of 1 $M_{\oplus}$
in the formation calculations.
This region is left empty because protoplanets inside the ice line quickly accrete all planetesimals
in their feeding zones.
Therefore, their luminosities are low
and are mostly attributed to gas accretion.
For a fixed core mass under these circumstances,
the envelope mass increases with orbital distance \citep[see][]{Ikoma2012}
because more gas can be bound at lower nebula temperatures,
which translates into a larger radius and makes
the hollow higher at larger distances (to approximately 2 AU).
Outside of this distance, another effect becomes dominant:
the solid accretion timescale is longer, which makes that
certain planets have high luminosities (due to planetesimal accretion)
at the end of the disk lifetime.
These planets can only hold tenuous envelopes
and no longer retain a relationship with the orbital distance.
In reality, there is no minimal core mass; therefore,
this empty region should not exist in the actual $a$-$R$ plot.
A very similar artifact can be seen in Fig. 7 of \citet{Owen2013}.
One real visible effect of planetary evolution can be observed by the decrease of the planetary radii of the entire population due to planet cooling.
Note the sharp upper limit for the planetary radius, which
is artificially sharp because, first, no special bloating mechanisms
are included \citep[e.g., ohmic heating,][]{Batygin2011},
and second, the opacity, which can affect cooling \citep{Vazan2013},
is identical for each planet.

The features related to evaporation are shown by the color of each point,
which represents the fraction of the initial envelope
that evaporated.
Here, we use black points to indicate the planets that
have lost their entire envelopes (the bottom left region of each panel).
As indicated, when a planet becomes a bare rocky core,
it settles to the bottom of the $a$-$R$ plane and detaches
from the planets that retain at least a portion of their initial envelopes;
this settling leads to the formation of an empty region that runs diagonally downward
in the $a$-$R$ plane between 0.06 and 0.5 AU.
This empty diagonal belt separates the bare rocky cores from the
planets that retain an envelope,
which, henceforth, we refer to as an ``evaporation valley".
The evaporation valley occurs because the radius of
a purely rocky planet is substantially smaller than that of a planet
with both a core and gaseous envelope.
For example, an envelope at only 0.1\% of the planetary mass
can dramatically enhance the planetary radius (Figure \ref{fig:mrRg}).
Additionally, the last 0.1\% of the envelope is lost on a
short timescale, $\sim 10^{5}$ yrs; therefore, we are unlikely to detect
a planet when it lies in the evaporation valley.
Thus, an empty valley appears in the $a$-$R$ plane after
many low-mass planets have become bare, rocky cores.
At 0.02 Gyr, the valley only appears within $\sim$ 0.1 AU,
and rapidly extends to $\sim$ 0.3 AU at 0.11 Gyr.
Clearly, the empty valley is only expected if all
low-mass planets begin with significant H/He envelopes, which
is unlikely. We discuss this topic further in $\S$ \ref{sect:dis41}.
The separated distribution of low-mass planets within 0.5 AU
suggests a bimodal size distribution of the close-in low-mass planets,
which was first theoretically observed by \citet{Owen2013}.
A similar but weaker structure was also demonstrated by \citet{Lopez2013}.
We show the size distributions for our
synthetic planet populations in $\S$ \ref{sect:comparison}.

\subsection{Synthetic Planets: B, influence of Parameters}

\label{sect:parameters}

To determine how our results depend on the evaporation description,
we simulate the evolution of the same nominal planet population but using different evaporation models.
Table \ref{tab:simulist} lists the details for these simulations.
Simulation XE is our reference.
Simulation NoEV is planetary evolution without evaporation.
Simulation SatE includes only EUV-driven evaporation and assumes that
the stellar EUV emissions are saturated during the first 100 Myr of planetary evolution.
Simulation XE2 uses the nominal evaporation model, but the heating efficiency
in both the X-ray ($\epsilon = 0.2$) and energy-limited EUV regime
($\epsilon = 0.12$) are twice as high compared with the XE simulation.
Simulation L12 uses the energy-limited model from \citet{Lopez2012}, which
uses the total flux between 1-1200 ${\rm \AA}$ as the incoming energy
and 0.1 as the heating efficiency.
Simulation B04 uses the energy-limited model from \citet{Baraffe2004},
which includes a different temporal evolution of the XUV emission from
a sun-like star and adopts a 100\% heating efficiency.
All the simulations are evolved for 10 Gyr.

\begin{figure*}
\includegraphics[width=18.0cm]{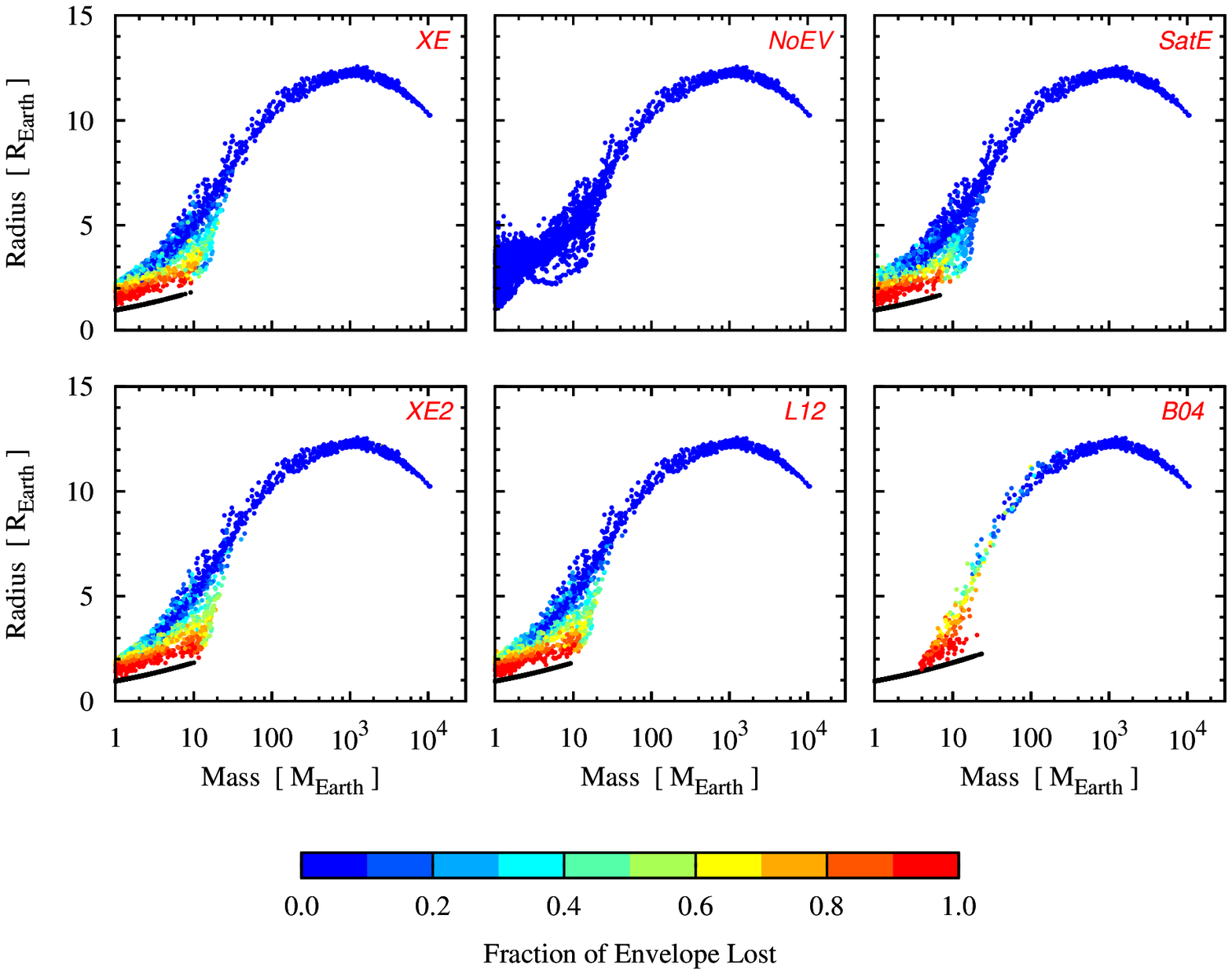}
\caption{
The mass-radius relationship of the planets between
0.06 and 1 AU at 5 Gyr for different simulations (Table \ref{tab:simulist}).
The color of each point shows how much of the initial envelope was lost.
The black points are planets that have lost all their initial envelopes.
All the planets have an identical, solar composition opacity,
a pure H/He envelope, and no bloating mechanisms are included.
This means that the vertical width of the  mass-radius relation is
likely underestimated.
The mass of the host star is 1 $M_{\odot}$.
All rocky planetary cores have a terrestrial composition (2:1 silicate-iron ratio).
}
\label{fig:6mr}
\end{figure*}

Because we are more interested in the potentially observable influence
of these different evaporation models, in Figure
\ref{fig:6arhoMdot}, we plot the mass-loss rates of the six simulations at 5
Gyr in the semi-major axis versus planetary mass $\times$ mean density
($M_{\rm p}^{2}/R_{\rm p}^{3}$) plane. At 5 Gyr, only
Simulation B04 leads to planets with evaporation rates
greater than $10^{-8}$ $M_{\oplus}$ yr$^{-1}$. Simulation B04 is
also the only simulation in which even a Jovian planet can evaporate
to a rocky core. At 110 Myr, the largest mass-loss rates of the gas
giants within 0.1 AU for Simulation B04 are $\sim$ $10^{-6}$
$M_{\oplus}$ yr$^{-1}$ (not shown in the plot), which indicates that these
planets lose at least 100 $M_{\oplus}$ in the first 100 Myr
of planetary evolution because the mass-loss rate of a planet decreases with time. Thus, many planets in Simulation B04, including certain close-in Jovian
planets, eventually lose their entire envelope. The mass-loss
rates for the other four simulations are significantly smaller; the
typical rate for a planet within 0.1 AU at 5 Gyr is $\sim$
$10^{-10}$ $M_{\oplus}$ yr$^{-1}$ and is $\sim$
$10^{-8}$ $M_{\oplus}$ yr$^{-1}$ at 110 Myr (not shown in the plot).

The evaporation timescale $M_{\rm p}/\dot{M}$ is
proportional to $M_{\rm p}^{2}/(R_{\rm p}^{3}F)$ ($F$ is the
incoming flux) in an energy-limited regime; thus, an evaporation threshold is
expected in the semi-major axis versus planetary mass $\times$ the mean
density ($M_{\rm p}\cdot\bar{\rho}$) plane
\citep{Jackson2012,Lopez2012,Owen2013}. Low-mass, low-density planets
below this threshold at the beginning of planetary
evolution will be evaporated to bare, rocky cores at high mean
densities, and eventually, no planets with H/He are below the
threshold, as demonstrated by the solid lines and black dots in Figure
\ref{fig:6arhoMdot}. 
A similar threshold is also apparent in
the NoEV simulation, which does not include evaporation; however, in this instance, the cut-off is unclear (as in the evaporation-inclusive simulations), and the threshold is at the lower value of $M_{\rm
p}\cdot\bar{\rho}$. 
This lower limit is an artifact of using a the minimal planetary
mass of 1 $M_{\oplus}$, which , indicates that the envelope
mass for a fixed core mass increases with orbital distance, as
discussed above. Evaporation renders this
threshold clearer and raises it in the $M_{\rm
p}\cdot\bar{\rho}$ plane. With a substantially stronger evaporation model,
the evaporation threshold is so high that it intersects the
bare, rocky cores; hence, the evaporation threshold becomes
blurred, as demonstrated in the B04 simulation. If planets with masses lower
than 1 $M_{\oplus}$ are included in our synthetic population,
then the evaporation threshold can also be blurred because the
low-mass bare rocky cores would extend to lower values of $M_{\rm
p}\cdot\bar{\rho}$ below the threshold. Examples of such low-mass
cores include $Kepler$-10b and Corot-6b (see Figure 6 in
\citet{Lopez2012}).

\begin{figure*}
\includegraphics[width=18.0cm]{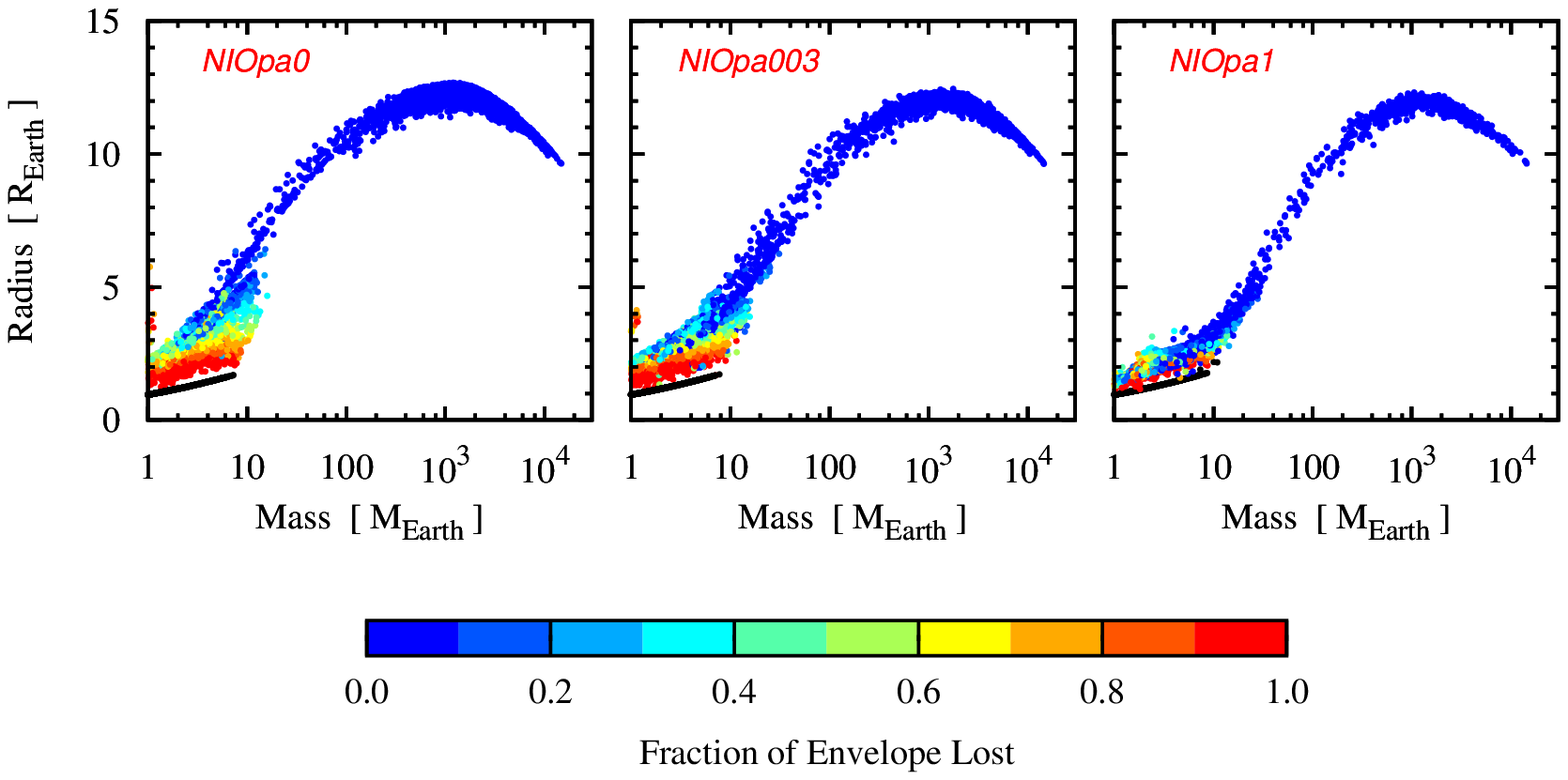}
\caption{
The mass-radius relationship of the planets between 0.06 and 1 AU 
at 5 Gyr of the three planetary populations
calculated using non-isothermal type I migration rates and
different grain opacities (Table \ref{tab:simulist}).
The color of each point shows how much of the initial envelope was lost.
The black points show the planets that have lost all their envelopes.
All of the planets  orbit around a sun-like star.
}
\label{fig:nio}
\end{figure*}

Figure \ref{fig:6aM} shows the $a$-$M$ distributions at 5 Gyr.
With the exception of Simulation B04, the remaining five simulations present
similar $a$-$M$ distributions.
The difference for B04 is the severe depletion of
planets with masses at 30-50 $M_{\oplus}$ within 0.2 AU.
At distances smaller than those modeled here ($a \lesssim$ 0.03 AU),
the observed planet population includes a desert that may be related to evaporation
\citep{Beauge2013,Kurokawa2014}.
In each simulation, the black bottom left corner corresponds to
the planets that have lost their entire envelope.
Although the heating efficiency in XE2 is twice that of XE, the black corner in XE2 does not show
a large difference compared with XE.
These data indicate that long-term evolution of the entire planet population
is not extremely sensitive to the heating efficiency of the evaporation model,
at least within the framework of the models listed here.
However, if a substantially more violent evaporation model is used, as for B04,
the $a$-$M$ distribution of the entire planet population noticeably changes.

Figure \ref{fig:6aR} shows the $a$-$R$ distributions of
the six simulations at 10 Gyr.
The color of each point indicates the ratio of
the envelope mass to the core mass, $f_{\rm env/core}$.
Here, all of the simulations that include evaporation show a
distinct feature compared with NoEV:
an evaporation valley in the radius distribution at $\sim$ 2 $R_{\oplus}$.
The $f_{\rm env/core}$ of each planet does not change in the NoEV simulation;
thus, based on the colors in the NoEV snapshot,
the initial $f_{\rm env/core}$ of a planet generally scales with its core mass.
This finding is expected based on the long Kelvin-Helmholtz timescales for gas accretion of low-mass cores.
For example, only the planets within 0.1 AU with a core larger than 10 $M_{\oplus}$
can have a $f_{\rm env/core}$ that exceeds 80\%;
low-mass planets with a core smaller than 2 $M_{\oplus}$ typically
have a $f_{\rm env/core}$ that is less than 10\%.
Thus, most low-mass planets with cores at
1-3 $M_{\oplus}$ can be quickly evaporated to bare, rocky cores
after evolution has begun (even in simulation SatE,
which includes the weakest evaporation model).
However, this clear valley only occurs if all these low-mass planets
begin with a primordial H/He envelope, which is
unlikely in reality.
In the actual formation process, some low-mass planets may reach their final mass only after the 
dissipation of the gaseous nebula, 
leading to planets without significant primordial H/He envelopes.
This is in contrast with
our synthetic population where no growth via giant impact
is included after the protoplanetary disk has dissipated.
When the low-mass planets become bare, rocky cores,
they settle to the bottom of the $a$-$R$ panel and are
separated from the planets that retain some H/He.
The evaporation valley becomes a large void region for B04,
for which the mass-loss rates are so high that most planets within 0.2 AU have evaporated to bare cores,
including certain gas giants with $M_{\rm p} \lesssim 400 M_{\oplus}$.
Such strong planet depletion between $\sim$ 2 and $\sim$ 10 $R_{\oplus}$
is not presented in the observed data (including $Kepler$ candidates), which
implies that an 100\%-efficient, energy-limited evaporation model
is incompatible with the observed radius distribution of the extrasolar planets.

\subsection{Mass-Radius Relationship of Close-in Planets}

\label{sect:mr}

Figure \ref{fig:6mr} compares the mass-radius relationship
of the planets between 0.06-1 AU at 5 Gyr
for the six simulations using different evaporation models.
The planetary radii plotted in the figure are
at the optical depth $\tau = 2/3$.
Compared with the old grey model \citep{Mordasini2012a}, the semi-grey model used
for the atmosphere increases the radius of close-in planets
due to stellar irradiation \citep{Guillot2010}.
This effect is demonstrated by the low-mass
planets at small semi-major axes,
such as planets with masses of 1-2 $M_{\oplus}$ but
radii of 3-5 $R_{\oplus}$ in the simulation NoEV, which
does not include an evaporation model.
These low-mass planets have low densities.
For example, the mean density of a 1.5 $M_{\oplus}$,
4 $R_{\oplus}$ planet is only $\sim$ 0.13 g ${\rm\,cm}^{-3}$.

The evaporation valley in Figure \ref{fig:6aR} and evaporation
threshold in Figure \ref{fig:6arhoMdot} can also be observed in the
mass-radius relationships in Figure \ref{fig:6mr}. The black bar at
the bottom of each panel in the five simulations that included evaporation
corresponds to the rocky cores of the planets that
have lost their entire envelopes.
Note that with the migration model used here,
all close-in low-mass planets that become bare cores
have only accreted inside of the iceline,
giving them a rocky interior. 
This could be different for higher migration rates,
or if several planets form concurrently in one disk
\citep{Alibert2013}.
In the current model, we assume that all
rocky cores have an identical composition (2:1 silicate:iron ratio,
as in Earth). In reality, this ratio (and the composition of
refractory element) depends on the stellar composition and a
planet's formation history, such as large impacts. The black bar
creates a gap at $\sim$ 2 $R_{\oplus}$, which separates the bare
cores from the planets that retain at least a portion of their
initial envelopes. This feature clearly corresponds to
the evaporation valley in Figure \ref{fig:6aR}). The length of this
black bar is related to the efficiency of evaporation mechanism. Simulation
SatE includes the lowest mass-loss rates; consequently, it produces the
shortest black bar. Simulation B04 produces the longest black bar,
extending to $\sim 20 M_{\oplus}$, and certain black points are
massive cores from stripped gas giants. In Simulation B04,
no planets with $M_{\rm p} \lesssim 5 M_{\oplus}$ within 1 AU maintain their primordial H/He. 
Another difference between NoEV and
the other five simulations is the disappearance of low-mass, very
low-density planets. This result is similar to the evaporation threshold in
Figure \ref{fig:6arhoMdot}. When evaporation is included in
planetary evolution, the bloated envelopes of close-in, low-mass
planets are rapidly removed. Thus, an upper
threshold is depicted in the bottom left corner of the mass-radius distribution
(for planets within 1 AU). Planets that are initially above this
threshold will lose at least a portion of their initial envelopes until
they are sufficiently dense that they lie below the threshold.
This threshold is also related to the efficiency of evaporation mechanism;
in Simulation B04, the threshold occurs at larger planetary
masses compared with Simulation SatE.

\subsection{Synthetic Planets: C, non-isothermal Migration and Different Grain Opacities}

\label{sect:nonnorminal}

Various evaporation models have been applied to the nominal
planet population that is calculated using the isothermal type I migration rate
reduced by a factor of 0.1 \citet{Tanaka2002}.
This is an artificial factor that prevents most synthetic planets
from falling into the star \citep{Benz2014}.
Recent studies have shown that, depending on the temperature profile of the disc,
type I migration can also induce outward migration \citep[e.g.,][]{Masset2006,Paardekooper2008,Kley2009,Uribe2011}.
Thus, in principle, the artificial reduction factor may be eliminated
even if the migration still seems to be too rapidly inward,
mainly due to saturation of the corrotation torques \citep{Benz2014}.
Here, we apply the nominal evaporation model to three synthetic populations that are all calculated using the full non-isothermal type I migration rate
from \citet{Dittkrist2014}.
The difference between these three populations is the ISM grain opacity
reduction factor, $f_{\rm opa}$, which is used during the formation stage.
For the planet population in Simulation NIOpa003, we use an $f_{\rm opa}$
at 0.003 (the nominal value) during formation;
in Simulation NIOpa0, the $f_{\rm opa}$ equals 0 (no grain opacity),
and in Simulation NIOpa1, the $f_{\rm opa}$ equals 1 (full ISM grain opacity).
The details are in Table \ref{tab:simulist} and \citet{Mordasini2014a}.
Lower grain opacity during formation phase yields a higher envelope mass for
a given core mass because it is more efficient to
radiate away the liberated potential energy,
allowing the envelope to contract.
This means that, at low $f_{\rm opa}$, planets with a lower mean density
are formed, which is indicated by
a larger maximal radius for a given mass in the mass-radius diagram.
In contrast, the atmospheric opacity during planetary evolution is identical
for all three populations and is given via the opacity of a grain-free gas
with a solar composition \citep{Freedman2008}.

The features that are related to evaporation, such as the black bottom left corner
in the $a$-$M$ diagram, the evaporation valley in the $a$-$R$ distribution,
and the black bar, which indicates purely rocky cores in the mass-radius plot,
have been detailed above
and are similar among these three populations.
Here, we focus on the features related to the effect of $f_{\rm opa}$.
Figure \ref{fig:nio} shows the mass-radius relationships of
the planets within 1 AU for the NIOpa003, NIOpa0, and NIOpa1 simulations.
One of the effects of the different grain opacities during formation
is the number of giant planets \citep{Mordasini2014a}.
As shown in the figure, the NIOpa0 group includes the most gas giants, while the NIOpa1 group shows the opposite result.
This effect is more clearly demonstrated by the histogram of
planet size distributions in Figure \ref{fig:9histo} in the following subsection.
Another effect is that, in Simulation NIOpa0, the largest overall radius
(at a mass of $\sim$ 4 $M_{\rm Jupiter}$) is slightly larger than for the other two simulations.
At $f_{\rm opa} = 0$, even very low-mass cores can become
supercritical and trigger giant planet formation \citep{Movshovitz2010,Mordasini2014a}.
These giant planets with low-mass cores 
yield large planetary radii.
A third effect, as extensively discussed in \citet{Mordasini2014a}, is
the effect of $f_{\rm opa}$ on the mass-radius relationship of low-mass planets.
With a low $f_{\rm opa}$, even cores with only a few $M_{\oplus}$ can accrete
significant envelope masses, thereby producing large radii.
This result is clearly visible in Figure \ref{fig:nio}.
For example, at 10 $M_{\oplus}$, the maximum radii are approximately 3.5, 5.5,
and 7.5 $R_{\oplus}$ for $f_{\rm opa}$ =1, 0.03, and 0, respectively.
The effect of the grain opacity during formation on the observable
mass-radius relationship as discussed in \citet{Mordasini2014a} (who neglected evaporation) is also found here, 
even if the very low-mass,
low-density planets are removed by evaporation.

\begin{figure*}
\includegraphics[width=18.0cm]{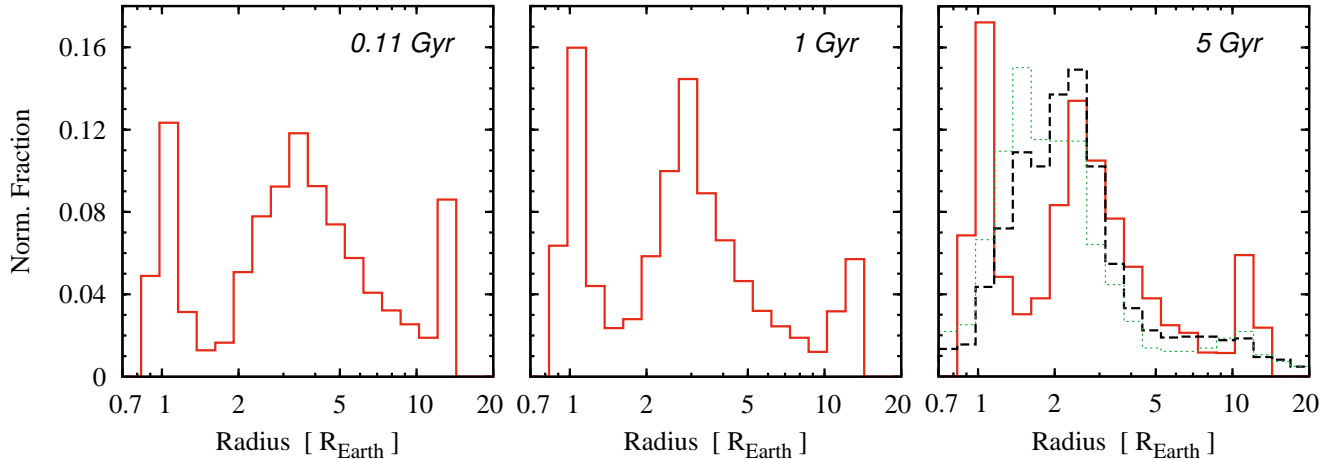}
\caption{
The normalized planet size distribution of
the XE simulation at 0.1, 1, and 5 Gyr.
The red solid line corresponds to the sub-population
of the planets within 1 AU.
In the 5 Gyr panel, the black dashed line shows the
size distribution of all $Kepler$ candidates,
while the green doted line shows the size distribution of
$Kepler$ candidates within 0.1 AU.
Note that at radii $\lesssim 2 R_{\oplus}$, the $Kepler$ data is
affected by observational bias, and the size distribution of
$Kepler$ candidates with correction for survey completeness is
a plateau at 1-3 $R_{\oplus}$ \citep{Dong2013,Petigura2013a}.
}
\label{fig:3histo}
\end{figure*}

\begin{figure*}
\includegraphics[width=18.0cm]{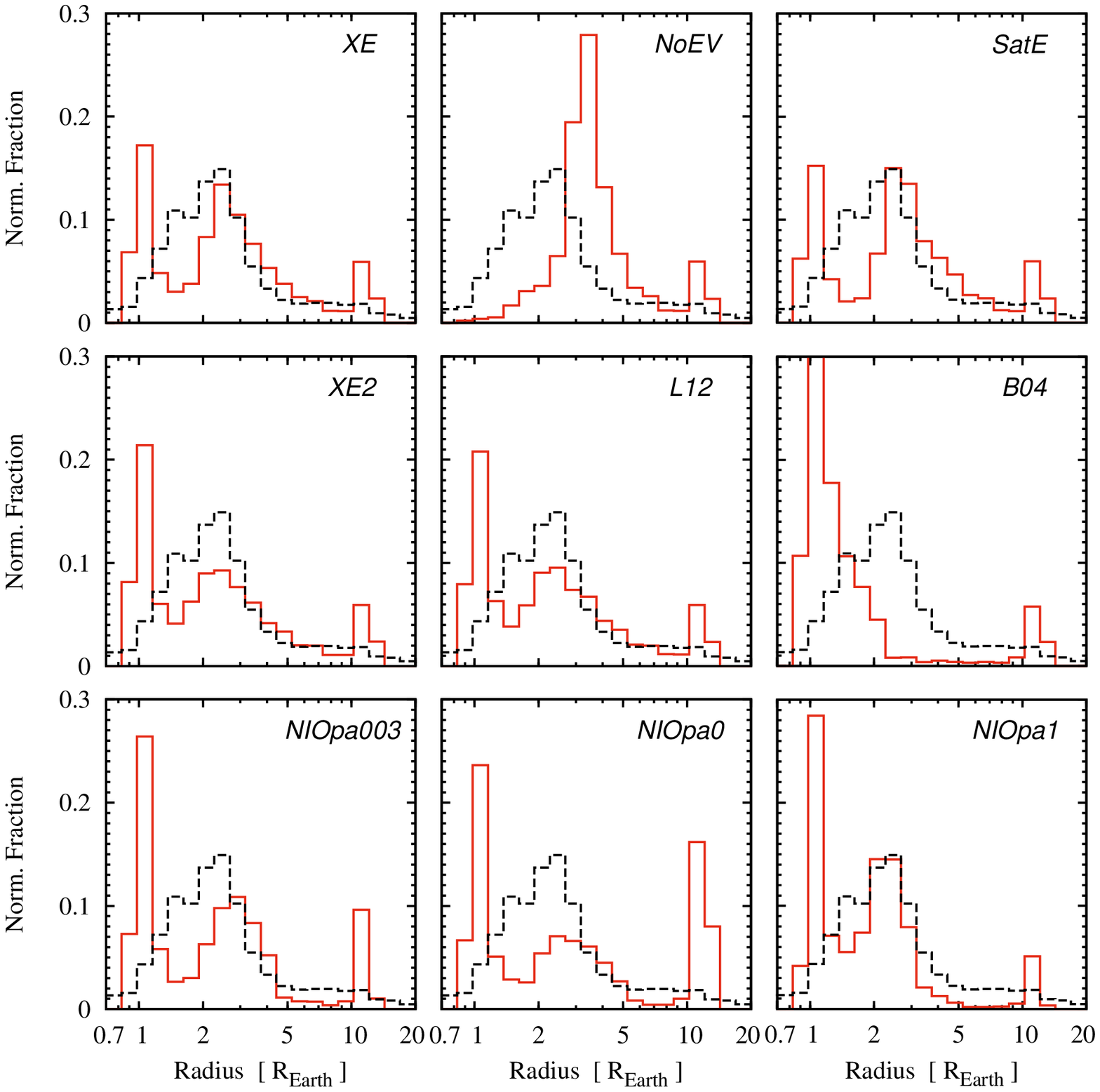}
\caption{
The normalized planet size distributions of all the simulations
at 5 Gyr (Table \ref{tab:simulist}).
In each panel, the red solid line shows the
synthetic population with 0.06 $< a$/AU $<$ 1,
and the black dash line shows size distribution of all $Kepler$ candidates.
}
\label{fig:9histo}
\end{figure*}

\subsection{Comparisons with $Kepler$ Candidates}

\label{sect:comparison}

Figure \ref{fig:3histo} shows the temporal evolution of the
normalized size distribution for Simulation XE.
For comparison with the observations, we only show the size distributions
for the planets within 1 AU.
There are three peaks in the planet size distributions,
all of which form at an early evolutionary stage.
The first peak is at $\sim$ 1 $R_{\oplus}$, which corresponds to the bare, rocky cores
of the low-mass planets that have entirely lost their initial envelopes.
The third peak, at 11 - 12 $R_{\oplus}$, $\sim 1 R_{\rm Jupiter}$, indicates
the sub-population of Jovian planets and is
a consequence of degeneracy of the electrons in the interior;
all giant planets with masses larger than $\sim$ Saturn have
the same radius of approximately $\sim 1 R_{\rm Jupiter}$.
Compared with the actual population, this effect must be overestimated in
our results because all planets evolved with the same opacity,
and bloating mechanisms are not included.
The middle peak at 2-4 $R_{\oplus}$ corresponds to the sub-population of
super-Earths and Neptunian planets that retain an envelope.
\citet{Owen2013} found that evaporation leads to a bimodal distribution
in planetary size with a planet deficit at approximately 2 $R_{\oplus}$.
Our results show a similar bimodal distribution with a minimum at
approximately 2 $R_{\oplus}$; however, this deficit is substantially more severe at $\sim$ 1.2-2 $R_{\oplus}$.

The normalized size distribution of all $Kepler$ candidates
(released Feb 26, 2012), most of which are within 1 AU,
are also plotted for comparison in the 5 Gyr panel.
Notably, the $Kepler$ data are biased and incomplete for $R \lesssim 2 R{\oplus}$
because the detection efficiency of the star and the detection sensitivity of the planet
decrease towards small planetary radii and larger semi-major axes.
The planetary occurrence of $Kepler$ candidates with a correction for
observational bias produces a nearly flat distribution at 1-3 $R_{\oplus}$
\citep{Howard2012,Dong2013,Petigura2013a,Petigura2013b}.
Figure \ref{fig:3histo} shows that the wide evaporation valley
in our result is not compatible with $Kepler$ data.
There are three possible reasons for this observation.
(1) Our evaporation model either overestimates evaporation or is too
deterministic because all stars in our simulation have the same mass and XUV flux as a function of time.
(2) We assume that all low-mass planets begin
with a primordial H/He envelope.
In reality, there is likely a population of
close-in, low-mass planets that formed without H/He via
planet impacts after the dissipation of the nebula 
\citep{Terquem2007,Ormel2012}.
(3) The distinct evaporation valley is related to our identical
core composition (2:1 silicate:iron ratio).
As mentioned, the migration model \citep[0.1 isothermal migration rate of][]{Tanaka2002}
predicts that all close-in low-mass planets loosing the entire envelope have a rocky interior (migration only inside of the iceline).
In reality, the actual core composition
of close-in planets might be highly diverse (i.e., contains some ice).
In the 5 Gyr panel, we also plotted the normalized size distribution
for $Kepler$ candidates within 0.1 AU,
which are planets that are more likely to be eroded by evaporation.
These candidates accumulate at $\sim$ 1-3 $R_{\oplus}$,
while the close-in $Kepler$ candidates
corrected for the observational bias have 
a nearly flat distribution at 1-3 $R_{\oplus}$
\citep{Dong2013,Petigura2013a,Petigura2013b}.
Thus, the evaporation valley in our synthetic planet populations
was not observed in the $Kepler$ data.
In a following paper, we show that, by varying the ice fraction of the
planetary cores, the radii of the bare, rocky cores can be substantially larger;
the dip in the evaporation valley is thus eliminated.
This means that important observational constrains for planet formation and migration theory can be 
deduced from evolutionary models with atmospheric escape \citep[cf.][]{Lopez2013}.

Figure \ref{fig:9histo} compares the normalized size distribution of
$Kepler$ candidates with our non-nominal simulations.
With the exception of NoEV, which does not include evaporation, and
B04, which has a 100\% heating efficiency in the energy-limited regime,
the remaining seven simulations show a bimodal distribution at small sizes.
As mentioned, an additional peak is at 1 $R_{\rm Jupiter}$ for the gas giants.
Simulation NoEV yields only one peak at 3-4 $R_{\oplus}$; this occurs
because the low-mass planets maintain their envelopes and, hence,
have large planetary radii.
The simulation B04 group is another extreme case that
yields a single peak at 1-2 $R_{\oplus}$.
Most of the low-mass planets in Simulation B04
evaporated to bare, rocky cores; thus, the number of Neptunian planets
that retain an envelope is too small to produce a second peak.
In the other seven simulations, the inner peak in the
bimodal distribution is at 1.0-1.2 $R_{\oplus}$, which
corresponds to bare, rocky cores with a 2:1 silicate:iron ratio.
For different simulations, the outer peaks in the bimodal distribution
differ slightly in both magnitude and location.
A stronger evaporation model will produce a lower outer peak
and move the position of the peak to smaller planet sizes,
as shown in the top two rows of Figure \ref{fig:9histo}, where
the only difference between these simulations is the evaporation model.
The effect of the grain opacity reduction factor, $f_{\rm opa}$,
is apparent in the bottom row of Figure \ref{fig:9histo}.
Because planets grow faster at smaller $f_{\rm opa}$,
the NIOpa0 group contains the largest number of gas giants.
Consequently, the NIOpa0 group has the highest peak at $\sim$ 12 $R_{\oplus}$; the NIOpa1 group is the opposite.

\begin{figure*}
\includegraphics[width=17.0cm]{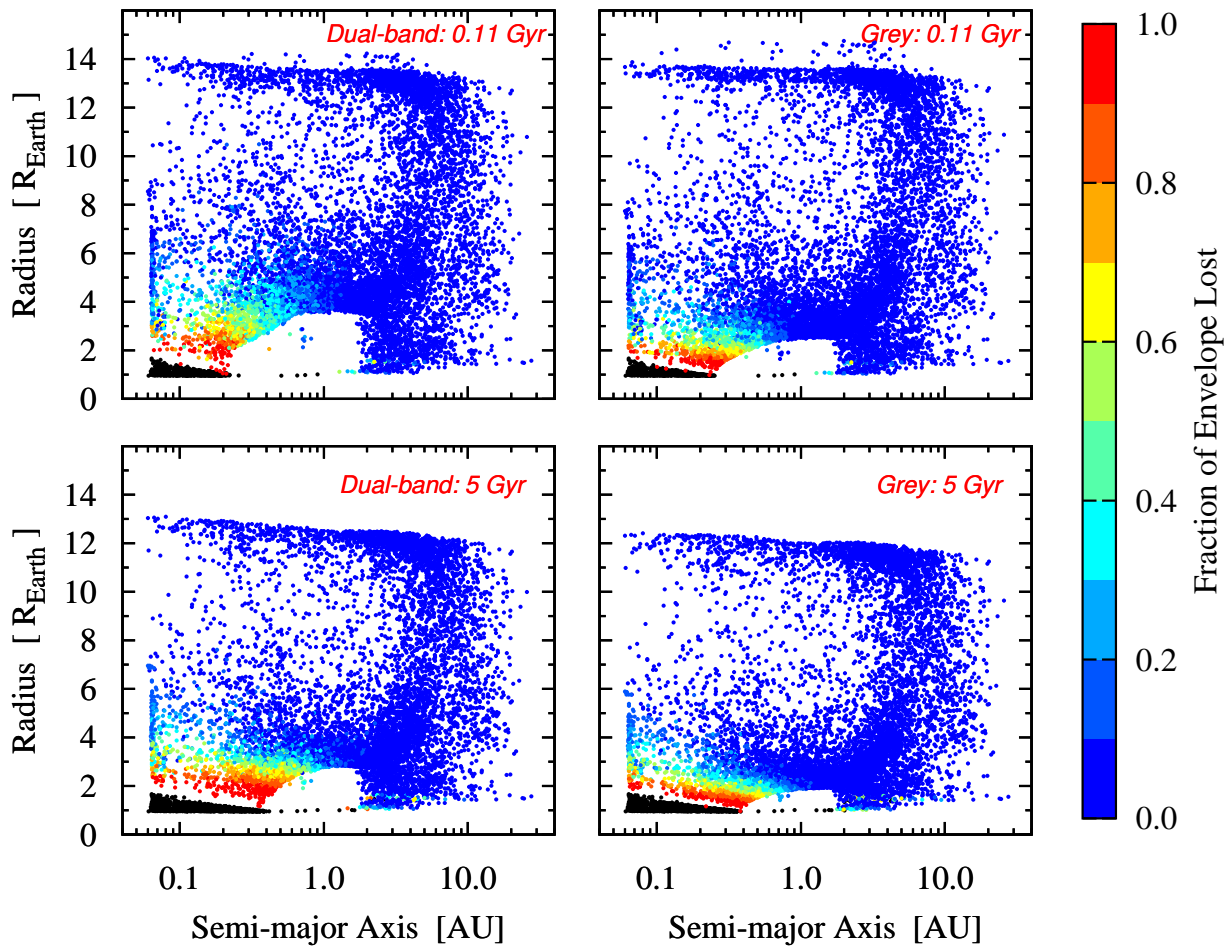}
\caption{
The $a$-$R$ distributions of the XE and GREY simulations at 0.11 and 5 Gyr.
The color of each point shows how much of the initial envelope was lost.
The black points are the planets that have lost all their initial envelopes.
The planetary radii are smaller and the evaporation valley is also narrower 
in the GREY simulation, 
but the general, population-wide impact of evaporation is
similar in the two simulations.
}
\label{fig:aRcomp}
\end{figure*}

\begin{figure}
\includegraphics[width=8cm]{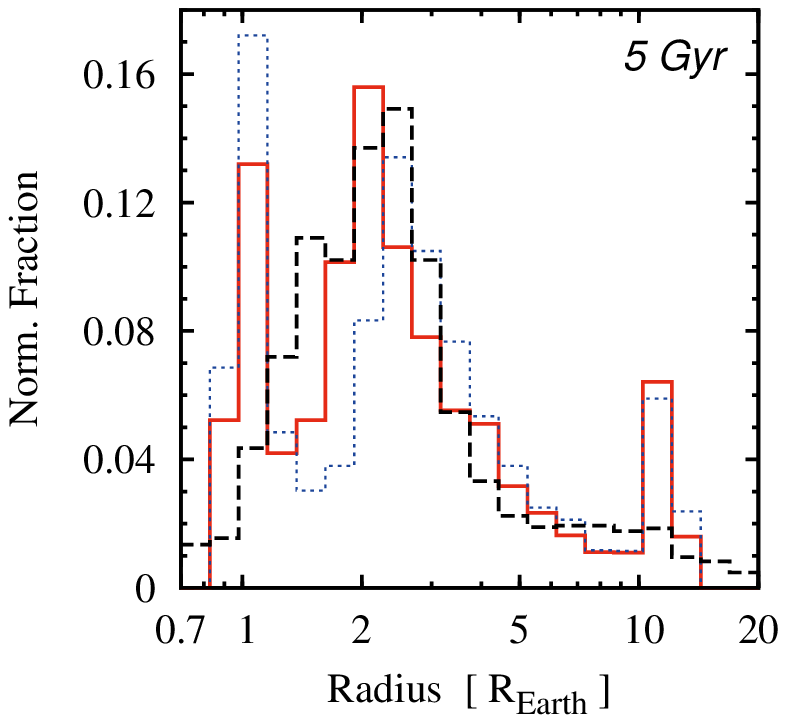}\\
\caption{
The size distributions of the planet
within 1 AU in the XE and GREY simulations at 5 Gyr.
The blue dotted line shows the XE simulation.
The red solid line shows the GREY simulation.
The black dashed line shows the size distribution of all $Kepler$ candidates.
}
\label{fig:1histo}
\end{figure}

\section{Discussion}
\label{sect:discussion}

\subsection{The Bimodal Distribution and Evaporation Valley}

\label{sect:dis41}

The bimodal distribution for planet sizes at approximately 2 $R_{\oplus}$
was first observed by \citet{Owen2013}, in which the
hydrodynamic evaporation of a theoretical planet population was studied.
\citet{Lopez2013} observe a diagonal band, on which planets
are relatively rare; the bimodal distribution
near $2 R_{\oplus}$ is less clear in their results.
In our results, this diagonal band \citet{Lopez2013}
becomes a distinct evaporation valley.
This evaporation valley separates the bare, rocky cores
from the planets that retain an envelope (Figure \ref{fig:6aR}).
The valley is $\sim$ 0.5 $R_{\oplus}$ wide and occurs at different
planet sizes (from $\sim$ 1 to $\sim$ 2.5 $R_{\oplus}$) depending on the semi-major axis.

We find that such empty valleys are closely related to the initial
characteristics of the synthetic planet population. In our synthetic
planet population, all close-in planets begin with a primordial H/He
envelope, and their rocky cores have a 2:1 silicate:iron ratio. Due to the long Kelvin-Helmholtz timescales
of low-mass cores, low-mass planets can only have a small initial
$f_{\rm env/core}$, which is proportional to the planetary core
mass. As shown in the NoEV simulation snapshot in Figure
\ref{fig:6aR}, low-mass planets within 0.5 AU with a radius $<$ 4
$R_{\oplus}$ typically include a $f_{\rm env/core}$ $<$ 10\%. Only
a planet with $>$ 6 $R_{\oplus}$ can have a $f_{\rm env/core}$
$\geq$ 80\%. Thus, low-mass planets are vulnerable to evaporation
to bare rocky cores due to their small envelopes and small
gravitational binding energies; these planets form a peak 
at about 1 $R_{\oplus}$ in the bimodal radius distribution of low-mass planets (trimodal, if the giants are included). 
Most close-in Neptunian planets can keep at least a portion of their
initial envelopes at the end of evolution; they form the second peak
at about 2-3 $R_{\oplus}$ in the bimodal distribution. In a forthcoming
paper, we show that the dip of the evaporation valley can be
removed by varying the ice fractions of planetary cores: the
sizes of bare low-mass icy cores can be $\sim$ 2 $R_{\oplus}$.

\subsection{Insensitivity to the Boundary Conditions}

Although the semi-grey model is a significant improvement
over the grey model used in \citet{Mordasini2012a},
it does not include the effects of
non-grey thermal opacities \citep{Parmentier2014m,Parmentier2013s}.
Neglecting non-grey effects in the planetary atmosphere boundary
may lead to an inaccurate planetary radius and, hence, mass-loss rates.
Fortunately, we find that the final radius distribution of the entire
planet population is not very sensitive to the outer boundary condition.
The exact planetary radius and mass-loss rate are clearly important
for individual planets but not so much for the overall statistical impact
on the planet population.

To clarify this, we perform a comparison numerical experiment,
the GREY simulation, using our nominal planet population but
with the previous grey atmospheric model from \citet{Mordasini2012a}.
The grey model assumes that stellar irradiation is absorbed
at the upper atmosphere where optical depth equals $2/3$.
For the GREY simulation, we also use the nominal evaporation model (i.e.,
both X-ray- and EUV-driven evaporation are included).
Figure \ref{fig:aRcomp} compares the $a$-$R$ distributions
of the GREY and XE simulations.
The planetary radii are smaller for the GREY model, as expected \citep{Guillot2010},
but the global effects of evaporation in general
and for the evaporation valley in particular
remain visible in a similar manner in the $a$-$R$ space.
The GREY simulation produces a narrower gap in the radius distribution
because the planetary envelopes from this numerical experiment are less bloated than
in simulations using the semi-grey model.
Figure \ref{fig:1histo} compares the planet size distributions
from the GREY and XE simulations at 5 Gyr.
The GREY simulation clearly shows a bimodal distribution at approximately 2 $R_{\oplus}$.
The difference is that, compared with XE,
the peak of the bare, rocky cores in the GREY simulation is less prominent,
and the outer peak in the bimodal distribution occurs at a smaller size.

\subsection{Simplifications in the Evaporation Models}

An accurate method to calculate the mass-loss rate due to
hydrodynamic escape is to solve the mass, energy,
and momentum conservation equations \citep{Tian2005,Penz2008,Murray-Clay2009,Owen2012,Lammer2013}.
In our work, we used an approximate
mass-loss formula in the energy-limited and radiation-recombination-limited regime as well as criteria for determining whether the outflow is EUV- or
X-ray-driven; we derived these from different authors \citep{Murray-Clay2009,Jackson2012,Owen2012}.
Thus, our evaporation model includes significant simplifications.
In reality, the evaporation efficiency that characterizes
the how much heating energy is converted to $P$d$V$ work 
depends on the characteristics of a planet 
and changes with time \citep{Yelle2004,Tian2005,Owen2012}.
Moreover, the direct transition from an X-ray regime to
an EUV-driven regime, as used in our model, can produce
a discontinuous change in the mass-loss rate,
as shown in Figure \ref{fig:single}.
Due to these limitations, we performed several groups of
population syntheses using different evaporation models, which allowed us to obtain a likely range of mass-loss rates.
Our results show that the evolution and final structure of
a single planet may be less accurate due to the evaporation model simplifications but that the statistical information
for the entire planet population was not greatly influenced.

For example, the XE2 simulation yields results
that are similar to those of the XE simulation even though the heating efficiencies
in the XE2 simulation are twice those of XE.
The final mass and radius distributions are not
sensitive to heating efficiency for the following reasons.
At a fixed incident flux, the evaporation threshold scales are
$M_{\rm env}M_{\rm p}/(\epsilon R_{\rm p}^{3})$.
As shown in the NoEV panel in Figure \ref{fig:6aR},
the initial $M_{\rm env}$ of a planet increases quickly
with the planetary mass.
For example, the initial $f_{\rm env/core}$ of a 4 $M_{\oplus}$
planet is typical of $\sim$ 10\%; however, for a
2 $M_{\oplus}$ planet, $\sim$ 1\% is typical.
For $R_{\rm p}$, the NoEV panel in Figure \ref{fig:6mr}
shows that the increase of the planetary radius 
with an increasing planetary mass is not significant
for low-mass planets with $M_{\rm p} < 10 M_{\oplus}$.
Thus, planetary mass plays a dominant role in determining the evaporation threshold.
Moreover, atmospheric evaporation is only intense in the
early stage for a short time (with a timescale $\sim$ 100 Myr).
When increasing the heating efficiency in an evaporation model, a slow increase is observed for
the critical planetary mass below which a planet will
be evaporated to a bare rocky core.
As shown in Figure \ref{fig:9histo}, the final planet size distributions
from the XE and XE2 simulations only show limited differences.
One exception is the B04 simulation, where even Jovian planets
lose a significant portion of their envelopes;
in this case, the bimodal feature at the small planet size disappears
because many Neptunian or Jovian planets evaporate
to bare, rocky cores. These massive cores finally
fill the deficit at $\sim$ 2 $R_{\oplus}$.

Another simplification in our models is that we
apply hydrodynamic evaporation to all planets,
even for those at large semi-major axes.
In fact, the atmospheric escape of planets at large distances
should be included in the Jeans escape regime.
Whether an escape flow is in the hydrodynamic or Jeans regime
can be assessed by comparing the mean free path of gas molecules
with the local scale height of the flow
(\citet{Johnson2013} demonstrate how to implement this criterion
in the energy-limited model).
\citet{Owen2012} show that, for close-in planets with semi-major axes
smaller than 0.1 AU, the dominant mass-loss process is hydrodynamic
evaporation. They found that at very small distances, such as $<$ 0.05 AU,
even gas giants with a few Jupiter masses can lose mass hydrodynamically.
They also show that massive planets with large densities are too
gravitationally bound for hydrodynamic outflow.
For example, a planet that is more massive than 1 $M_{\rm Jupiter}$
with a density greater than 1 g\,cm$^{-3}$ can no longer
undergo hydrodynamic evaporation at $\sim$ 1 AU.
Thus, our models overestimate the mass-loss rates
of the planets at large distances by assuming
that the planets undergo hydrodynamic evaporation.
However, this assumption does not affect the main statistical results for
the entire planet population because only
low-mass planets with tenuous envelopes can be evaporated
to bare cores at large distances due to the overestimated mass-loss rates.
We also did not include non-thermal ion escape in our
models (e.g., the planetary ions that are captured by stellar wind).
The influence of non-thermal escape on the statistical results is
also weak because the mass-loss rate of ion pick-up escape is typically
several times smaller than for thermal atmospheric escape
\citep{Kislyakova2013}.

\subsection{Dependence on Stellar Type}

Most $Kepler$ candidates orbit a host star
with a mass of 0.8-1.1 $M_{\odot}$. 
During the formation and evolution of our synthetic populations,
all planets orbit around a sun-like star with a mass of exactly 1 $M_{\odot}$.
The key effect of different stellar types is that they
lead to different disk properties, which finally determine the
characteristics of the synthetic planet populations \citep{Ida2005,Alibert2011}.
Massive F/G-type stars are found to yield fewer Neptunian planets
but more massive Jovian planets.
K/M-dwarfs are predicted to yield more Neptunian planets
but fewer Jovian planets \citep{Ida2005,Alibert2011}.
However, for the low-mass planets,
which are sensitive to evaporation
and form a bimodal size distribution,
the fraction does not strongly depend on the mass of the central star \citep{Ida2005,Alibert2011}.
Thus, at least for 0.8-1.1 $M_{\odot}$, comparisons between $Kepler$ candidates
and our synthetic planet populations are feasible.

The stars in our simulations all have the same $L_{\rm EUV}$ and $L_{\rm X}$,
which follows the temporal evolution of the X-rays and EUV emissions of
a sun-like star.
In reality,
the EUV flux (360-920 ${\rm \AA}$) of a sun-like star
at different times does not have direct observational constraints, with the exception of the present Sun.
The temporal evolution of the EUV flux adopted in this work is
from \citet{Ribas2005}, which was derived by scaling the EUV flux with
the temporal evolution of the stellar flux in other wavelength ranges.
The accuracy of this method is approximately 10\% - 20\% \citep{Ribas2005}.
Considering that the EUV evaporation contributes typically less than 10\% of
the total mass-loss of a planet \citep{Owen2013}, the low accuracy of
the EUV fluxes at different times does not affect the statistical results.
On the other hand, the stellar X-ray emissions are highly diverse \citep{Guedel2004}.
The ratio of X-rays to bolometric luminosity during the early saturated phase
decreases from $10^{-3.1}$ for late K dwarfs to $10^{-4.3}$
for early F-type stars (0.29 $\leq$ $(B - V)_{0} <$ 1.4) \citep{Jackson2012}.
Thus, our model is too deterministic because we use a fixed evolution track of
stellar X-ray emission.
M dwarfs can have strong chromosphere activities.
They are very bright regarding hard radiation, which
can strongly erode the planetary envelope.
However, planets around K/M-dwarfs also have a less effective
Roche lobe effect \citep{Lammer2009,PenzMicela2008}.
For planets surrounding F-, G-, K-, and M-type stars,
\citet{Lammer2009} show that evaporation can only remove
a limited portion of the initial gas giant envelopes,
and only Neptunian and terrestrial planets are significantly affected by evaporation.

\section{Summary}
\label{sect:summary}

We combine models of hydrodynamic atmospheric escape with planet population syntheses which include both planet formation and evolution. Our global planet formation model is constructed based on the core accretion paradigm. We find that atmospheric escape adds characteristic features to the radius distribution of the synthetic planetary populations. The most interesting imprints are an “evaporation valley” in the radius-distance diagram and a bimodal planet size distribution \citep[cf.][]{Owen2013}. These features are consequences of evaporation, but their properties are also related to the characteristics of the initial planet population, and thus the planet formation process.  

In our synthetic populations, the initial envelope fraction of low-mass planets with sizes of less than 4 $R_{\oplus}$ is normally $<$ 10\% (especially for planets at small distances). The initial mass fraction of planetary envelopes also scales with the planetary core mass and is typically $<$ 5\% for Earth-size planets with a 1 $M_{\oplus}$ core. Due to their low gravity, such very low-mass planets are sensitive to evaporation and are evaporated to bare cores with radii of about 1 $R_{\oplus}$ at sufficiently small orbital distances. However, planets with larger core masses also have larger envelopes, as typical for the core accretion paradigm. At the end of evolution, such more massive super-Earths or Neptunian planets retain at least a portion of their H/He envelopes. They will have substantially larger radii because  already 0.1\% in mass of H/He significantly increases the radius. The threshold core mass where a complete loss of the initial H/He envelope occurs decreases with orbital distance. 

As a result, an ``evaporation valley" running diagonally downward in the orbital distance - planetary radius plane appears. It separates bare cores from planets retaining some primordial H/He. As the process of losing the last, e.g., 0.1\% of H/He occurs on a short timescale, the valley is sparsely populated with planets at any given moment at late times (e.g., 5 Gyrs). At this time, the evaporation valley runs diagonally downward from about 2 $R_{\oplus}$ at 0.06 AU to 1 $R_{\oplus}$ at 0.5 AU.

Corresponding to this valley, the one-dimensional radius distribution of close-in low-mass planets is bimodal, with a local maximum at about 1 $R_{\oplus}$, a local minimum at about 1.5 $R_{\oplus}$ and another maximum at 2-3 $R_{\oplus}$. The lower maximum in the bimodal distribution corresponds to the bare cores of planets that have lost their entire initial H/He envelope. The minimum corresponds to the ``evaporation valley". The second maximum corresponds to low-mass planets that have kept some primordial H/He.

No such very prominent features (deep diagonal evaporation valley and strong depletion at about 1.5 $R_{\oplus}$ in the radius histogram) can be seen in the $Kepler$ data for $R_{\rm p} < 2 R_{\oplus}$, even if a small local minimum might be present also in the observational data \citep{Owen2013,Petigura2013a}. This difference could be due to the following reasons: 

-First, our evaporation model might be too deterministic (identical mean XUV flux as a function of time for all stars) and/or overestimate the impact of evaporation. 

-Second, in our formation model, all low-mass planets start with a primordial H/He envelope and reach their final mass during the presence of the gaseous disk. In reality, at least some terrestrial planets will reach their final mass only after the dissipation of the gaseous nebula through a series of giant impacts, as likely for the Earth itself. This population of ``late" planets is not expected to start with a significant H/He envelope (a few percent), and therefore will not exhibit the imprints of evaporation (no significant radius evolution in time). Such planets would fill in the valley. For such planets, the transition from solid planets to planets with H/He should likely not be a clear function of the semi-major axis, in contrast to the case that the transition is due to evaporation as studied here. This is an important constraint for formation models.

-Third, the composition of the bare cores might be different than in the model here: in our synthetic populations, the sizes of the bare evaporated cores range between 1.0 and 1.2 $R_{\oplus}$ because all of the low-mass planets within 1 AU have rocky cores with a zero ice fraction despite orbital migration included in the formation model: they only migrate within the iceline. If the migration of individual planets is more efficient than assumed here, or if more massive planets push lower-mass planets closer-in due to capture in mean motion resonance in multiple systems (neither included here), planets that have formed beyond the ice line may migrate to close-in orbits during the formation stage. Such planets will accrete high amounts of ice during formation and, hence, have a large, icy core. In a forthcoming paper, we use a synthetic planet population with icy planetary cores to show that the sizes of the bare icy cores will be significantly larger and that these bare icy cores can fill in the minimum at $\sim 1.5 R_{\oplus}$. Thus, if close-in planets have both rocky and icy (or mixed) cores, this can lead to an approximately flat radius distribution for $R_{\rm p} < 2-3 R_{\oplus}$, as observed in the bias-corrected radius distribution of the $Kepler$ sample \citep[e.g.,][]{Fressin2013,Petigura2013a}. Thus, a diversity in the core composition combined with the consequences of evaporation may provide an explanation for the observed radius plateau to 2-3 $R_{ops}$ and the decrease at larger radii.  Clearly, the ice content of close-in low-mass planet is another fundamental constraint for formation (migration) models.  

The specific shape and location of the second maximum at 2-3 $R_{\oplus}$ in the bimodal distribution (planets that have retained primordial H/He) is related to envelope evaporation. Stronger evaporation produces a lower outer maximum and moves the peak to smaller radii. Most of our evaporation models lead to a similar outer peak, which is approximately consistent with the size distribution of the $Kepler$ candidates in the radius range of about 2-8 $R_{\oplus}$. However, in two extreme cases, the NoEV simulation without any evaporation and the B04 simulation with very strong evaporation (100\% heating efficiency), the final planet size distribution shows clear differences compared with the $Kepler$ data in this range. This indicates that evaporation is indeed important in shaping the characteristics of close-in, low-mass planets \citep{Lopez2012,Owen2013}. For the comparison of observations with predictions of formation model for such planets, it should therefore be taken into account. Other major findings are as follows: We find that in contrast to the radius distribution, the mass distribution of the entire planet population is barely affected by evaporation at $a >$ 0.06 AU: low-mass planets may lose all H/He, but its initial mass fraction is low anyway. Giant planets in contrast do not lose much H/He in our nominal evaporation model - their high mass gravity protects them. 

We demonstrate the importance of the core mass using a parameter study that is similar to the study described in \citet{Lopez2012}. We confirm the evaporation threshold in the $M_{\rm p}^{2}/R_{\rm p}^{3}$ versus distance$^{2}$ plane \citep{Jackson2012,Lopez2012,Owen2013}. Furthermore, we find that this evaporation threshold is also apparent in the mass-radius relationship of close-in planets. For the simulations that include evaporation, the mass-radius relationship clearly contains a threshold, and the very low-mass, very low-density planets that are initially above the threshold will lose at least a portion of their envelopes until they are below this threshold. Finally, the impact of the grain opacity in the outer radiative zone of protoplanets during the formation stage on the mass-radius relationship at 5 Gyrs remains clear also with evaporation. 

Our study shows that several important observational constrains can be inferred from the comparison of observational results and theoretical formation and evolution models that include atmospheric escape. This is of high interest in view of several future high-precision photometric missions like TESS \citep{Ricker2010} or CHEOPS \citep{Broeg2013}. Our results in particular show a dynamical evolution of the planetary population in terms of the radii (or composition) in time. In principle, such a temporal evolution could be observed directly with PLATO 2.0 \citep{Rauer2013} which determines the ages of the host stars. This would open a new perspective to understand the nature of close-in planets.

\acknowledgements We thank Dr. Jonathan Fortney for the atmospheric
structures used for the comparison with the semi-grey model. We
also thank Dr. Helmut Lammer, Kai-Martin Dittkrist, and
Gabriel-Dominique Marleau for helpful discussions. S. Jin
acknowledges the financial support of the Chinese Academy of
Sciences and the Max-Planck-Gesellschaft. This work was also
supported by the National Natural Science Foundation of China
(Grants No. 11273068, 11473073), the Natural Science Foundation of Jiangsu
Province (Grant No.  BK20141509), the innovative and
interdisciplinary program by CAS (Grant No. KJZD-EW-Z001), and the
Foundation of Minor Planets of the Purple Mountain Observatory. J.
H. Ji acknowledges the Strategic Priority Research Program - The
Emergence of Cosmological Structures of the Chinese Academy of
Sciences (Grant No. XDB09000000). C. Mordasini thanks the
Max-Planck-Gesellschaft for the Reimar-L\"ust Fellowship.
We thank the referee for comments that helped to improve the manuscript.

\end{document}